\DeclareUnicodeCharacter{2212}{-}
\documentclass[twocolumn]{aa} 

\usepackage{sidecap}
\usepackage{xcolor}
\usepackage{listings}
\usepackage{natbib}
\usepackage{subcaption}
\bibpunct{(}{)}{;}{a}{}{,} 
\usepackage{booktabs,caption}
\usepackage[flushleft]{threeparttable}
\usepackage{adjustbox}

\usepackage{graphicx}
\usepackage{graphics}
\usepackage{txfonts}
\usepackage{xcolor}
\usepackage{listings}
\usepackage{natbib}
\usepackage{subcaption}
\usepackage{booktabs,caption}
\usepackage[flushleft]{threeparttable}
\usepackage{adjustbox}
\usepackage{hyperref}
\usepackage{float}
\usepackage{soul}

\begin{document}

   \title{A detailed star formation history of the distant Milky Way satellite Canes Venatici I: An old dwarf spheroidal forming most of its stars after reionisation}
   
   \author{P. Pérez-García\inst{1,2},
           C. Gallart\inst{2,1},
           M. Monelli\inst{3,1,2},
           S. Cassisi\inst{3,4},
           G. Battaglia\inst{2,1},
           L. Carigi\inst{6},
           E. Fernández-Alvar\inst{1,2},
           P. B. Stetson\inst{5},
           \and
           F. Surot\inst{2,1}}

   \institute{
   Departamento de Astrof\'isica, Universidad de La Laguna, E-38205 La Laguna, Tenerife, Spain
   \and 
   Instituto de Astrof\'isica de Canarias, Calle Vía Láctea s/n, E-38206 La Laguna, Tenerife, Spain
   \and
   INAF -- Osservatorio Astronomico di Abruzzo, via M. Maggini, sn, 64100 Teramo, Italy
   \and
   INFN, Sezione di Pisa, Largo Pontecorvo 3, 56127 Pisa, Italy
   \and
   Herzberg Astronomy and Astrophysics, National Research Council, 5071 West Saanich Road, Victoria, British Columbia V9E 2E7, Canada 
   \and
   Universidad Nacional Autónoma de México. Instituto de Astronomía. A.P. 70-264, 04510. Ciudad de México, México
   }

   \titlerunning{Detailed star formation history of dSph CVnI}
   \authorrunning{Pérez-García, P.J. et al.}

   \date{Received --, --; accepted --, --}

 
  \abstract
   {Canes Venatici I (CVnI) is a distant Milky Way (MW) dwarf spheroidal (dSph) satellite. To date, there is no consensus on its evolution and the properties of its stellar population.}
   {This work is aimed at obtaining an accurate picture of the star formation history (SFH) and metallicity distribution of CVnI to reconstruct its evolution in detail, as well as to infer the possible effects of internal and external feedback.}
   {We used archival HST/ACS observations to obtain the deepest colour-magnitude diagram (CMD) to date, reaching $\sim$2 mag below the old main sequence turn-off. We derived the SFH and metallicity distribution of CVnI using a state-of-the-art CMD-fitting technique.}
   {We found that the SFH is dominated by a single star formation period, which peaked around 11.8 Gyr ago, well after the reionisation epoch, and remained active until $\sim$10.4 Gyr ago. The galaxy's global metallicity, [Z/H], ranges from -2.4 to -1.5 dex. Interestingly, while its star population is old and metal-poor, compared to other MW satellites of a similar size and luminosity that formed most of their stars in a single burst, CVnI experienced a delayed star formation period. We carried out an exhaustive set of tests with synthetic and real populations to ensure the reliability of these results.}
   {We suggest that the evolution of CVnI could have been regulated by its mass accretion history. The galaxy would have evolved above (but still close to) a theoretic redshift-dependent critical mass above which it is possible to maintain star formation. The combined effect of SN feedback would have made CVnI's halo mass subcritical at z$\approx$2, definitively quenching its star formation. We cannot rule out the possible influence of gravitational interaction of CVnI with the MW at early ages, as hinted at by the most recent calculations of its orbit.}


   \maketitle
%

\section{Introduction}

Local Group (LG) dwarf galaxies can be considered as ‘fossils’ from the Universe's early stages. The study of the evolution of these stellar systems provides constraints on the physical processes at play in the early LG environment and also on possible interactions with the Milky Way (MW) or Andromeda. Due to the proximity of LG dwarfs, we can obtain high precision photometry and spectroscopy of their individual stars, which provides the opportunity to reconstruct the star formation history (SFH) and chemical evolution of the host system in the best possible detail.

For these reasons, the study of dwarf galaxies belonging to the LG has experienced a significant boom since the end of last century, supported by technological advances such as ground-based wide-field imagers, multi-object spectroscopy, and space missions (the Hubble Space Telescope \footnote{https://www.nasa.gov/mission\_pages/hubble/main/index.html} and Gaia \footnote{https://www.esa.int/Science\_Exploration/Space\_Science/Gaia}, notably). 

It was by the middle of the first decade of the 21st century when the new discoveries began to extend to the ultra-faint regime \citep[e.g.][]{2019ARA&A..57..375S}\footnote{In this work, Simon suggests that dwarf galaxies with absolute magnitudes fainter than $M_V$ = $−$7.7 ($L_V$ = 10$^{5}$ L$_{\odot}$) should be considered Ultra Faint Dwarfs (UFDs).}, revealing galaxies of just a few hundred or thousands of {M$_\odot$} in stellar mass \citep[e.g.][]{2005AJ....129.2692W, 2006ApJ...647L.111B}. The emergence of imaging surveys such as SDSS\footnote{https://www.sdss.org/}, Pan-STARRS \citep{2016AAS...22732407C}, DESI Legacy \citep{2019AJ....157..168D}, DELVE \citep{2021ApJS..256....2D}, and  PAndAS \citep{2009AAS...21330705M} have enabled the discovery of many tens of dwarf galaxy satellites of the MW and Andromeda \citep[see e.g. fig 1 in][]{2025OJAp....8E.142P}.
The population of LG galaxies is still steadily increasing. As an example, between 2020 and 2025 alone, new possible discoveries of dwarf galaxies in the LG, such as Eridanus IV \citep{2021ApJ...920L..44C}, Centaurus I and Delve I \citep{2020ApJ...890..136M}, Pegasus V/Andromeda XXXIV \citep{2022MNRAS.515L..72C}, Aquarius III \citep{2025ApJ...979..164C}, or Leo VI \citep{2025ApJ...979..176T}, were reported.

In parallel with this growing increase in the identification of new dwarf galaxies, the tools for their study have also undergone significant evolution. Theoretical stellar evolution libraries, such as PARSEC \citep{2000A&AS..141..371G}, Dartmouth \citep{2008ApJS..178...89D}, MIST \citep{2016ApJS..222....8D,2016ApJ...823..102C}, or the Bag of Stellar Tracks and Isochrones (BaSTI-IAC) project \citep{2018ApJ...856..125H,2021ApJ...908..102P} have been developed and periodically updated in the input physics. Likewise, the methods of analysis have also been developed and improved to extract the star formation rate (SFR) and the chemical evolution history (which we refer to globally as the SFH in this work) of a stellar system from deep colour-magnitude diagrams (CMDs); in particular, the CMD fitting technique \citep{1991AJ....102..951T,1992ApJ...388..400B,1996ApJ...462..672T,1997NewA....2..397D,2009ARA&A..47..371T,2010AdAst2010E...3C,1996AJ....112.1950G,1999AJ....118.2245G,2024A&A...687A.168G}. From the first sketches with limited resolution, deeper data coupled with improved algorithms currently provide SFRs and age-metallicity relations at a resolution better than 1 Gyr at the oldest ages. Moreover, multiplex spectrographs provide complementary abundance determinations for a large sample of stars. 

One aim of studying these dwarf galaxies is to clarify how their star formation histories (SFHs) have been shaped by processes regarded as key drivers of early-universe evolution, including the UV background, supernova feedback (SNe), and environmental effects associated with dense regions or with the vicinity of massive galaxies such as the MW. The mass of the galaxies will have acted as modulator of the relative impact that each of these processes will have had on their evolution, with those with the greatest mass, baryonic and dark, expected to be the least affected by them in terms of the duration of their star formation period.

Although hydrodynamic simulations \citep{2017MNRAS.471.4128P} have shown that SN feedback should greatly impact the evolution of dwarf satellite galaxies, it can also be expected that the progenitors of the smallest ones (e.g. ultra-faint dwarfs, UFDs) would also be intensely affected by reionisation and/or by the environment. \citet{2015ApJ...811L..18G} established two main classes of dwarf galaxies in the LG, fast and slow, based on the characteristics of their lifetime SFHs, and in particular, on the fraction of the old ($>$ 10 Gyr) population. They linked the origin of these two types primarily to the density of the formation environment, with fast dwarfs assembling early and quickly in high-density environments, where interactions triggering star formation were likely leading to high SFRs even before reionisation. Strong gas loss would follow as a consequence of reionisation and SN feedback acting together.
In turn, the impact of reionisation and UV feedback would also depend on the mass acquired by the galaxy before reionisation was complete \citep{2015MNRAS.450.4207B}. In this regard, the so-called 'primordial scenario' \citep[see][]{2009ApJ...693.1859B, 2009MNRAS.395L...6S} suggests that UFDs are pre-reionisation fossils formed before reionisation in dark halos with circular velocities < 20 km s$^{-1}$ (corresponding approximately to a mass $M$ < 10$^{8}$ M$_{\odot}$, which is typical for systems related to present-day UFDs), in which the UV feedback from massive stars would have played a fundamental role in suppressing the star formation before the end of the reionisation, while the heating of the gas by reionisation itself would have prevented the mass accretion necessary to continue star formation. In contrast, galaxies hosted in more massive dark matter halos could have maintained star formation after the end of reionisation \citep[e.g.][]{2010MNRAS.402.1599S, 2015MNRAS.453.1305W, 2019A&A...630A.140R}. In any case, this mass-dependent effect of reionisation on the suppression of star formation is not clearly established. For example, the LCID project \citep[and references herein]{2015ApJ...811L..18G}, which investigated a sample of isolated dwarf galaxies in the LG, has shown that reionisation alone did not stop star formation in galaxies with baryonic masses between 10$^{6}$ and 10$^{7}$ M$_{\odot}$ \citep{2010ApJ...722.1864M, 2010ApJ...720.1225M, 2011ApJ...730...14H, 2013ApJ...778..103H}. Meanwhile, other studies, such as that by \citet{2014ApJ...796...91B} carried out on six UFDs, supported the hypothesis that a global influence, such as reionisation, synchronised the complete suppression of star formation in these smaller systems. 

To further investigate the effects of these processes in the early universe, UFDs are key objects, which are proving to be more complex galaxies than initially thought. Mostly old and relatively chemically under-evolved, we would expect them all to share similar SFHs with suppressed star formation at very early ages and with very homogeneous metallicity populations. However, there is evidence of cases with prolonged star formation periods up to intermediate or even recent ages, such as LeoT \citep{2012ApJ...748...88W}, and large dispersions in metallicity, such as Bootes I \citep{2011ApJ...738...51L} or Reticulum II \citep{2016ApJ...830...93J, 2023ApJ...944...43S}.

Canes Venatici I (CVnI) is one of the dwarf satellite galaxies of the MW. It was discovered by \citet{2006ApJ...643L.103Z} as a stellar overdensity in the north Galactic cap using Sloan Digital Sky Survey Data Release 5 (SDSS DR5). They estimated an absolute magnitude $M_V$ $\sim$ −7.9 mag, a central surface brightness of $\mu$$_{0}$ $\sim$ 28 mag arcsec$^{-2}$, a distance of $\sim$ $220_{-16}^{+25}$ kpc and a half$-$light radius of $\sim$ 474 pc. It has normally been classified as an UFD, but \citet{2019ARA&A..57..375S} suggests that this could be due to its discovery at the same time as other fainter galaxies (probably due to its lower central surface brigthness and larger distance). However, its luminosity $M_{v}$ (-8.8) stands out from the fainter population, more so resembling those of classical dSphs, such as Ursa Minor. 

There is also general agreement that CVnI has a predominantly old (> 10 Gyr) and low-metallicity stellar population, namely, of [Fe/H] < $-$1.8 dex on average \citep{2006MNRAS.373L..70I,2007ApJ...670..313S,2008AJ....135.1361D,2008ApJ...674L..81K,2008ApJ...685L..43K,2011ApJ...727...78K,2011ApJ...727...79K,2008ApJ...672L..13M,2010MNRAS.402.1357U,2012ApJ...744...96O,2012AJ....144....4M,2013ApJ...767..134V,2014ApJ...789..147W,2014ApJ...789..148W,2018ApJ...860...66M,2020ApJS..247....7H,2023ApJS..268...19W}. However, there are discrepancies between works by different authors about the possible existence in CVnI of other significantly younger and/or more metal-rich populations. \citet{2006MNRAS.373L..70I} carried out a spectroscopic study and reported the existence of a group of stars with a radial velocity dispersion of $<$ 1.9 km s$^{-1}$, which they identified as a cold kinematic component related to a population of young stars with a metallicity range between ($-$1.9 $\leq$ [Fe/H] $\leq$ $-$1.5). \citet{2008ApJ...674L..81K} reported the possible existence of stars aged 5, 1.5, and 0.6 Gyr, all with metallicity [Fe/H] $\sim$ $-$2.0, based on the superposition of isochrones on the galaxy's CMD in (B, B-V). \citet{2008ApJ...672L..13M}, defended the existence of a young population of stars with ages $\sim$ 1.4 – 2.0 Gyr and metallicity [Fe/H] $\sim$ $-$1.5, identified in a blue plume (BP) positioned in the (V, B$-$V) CMD at B$-$V $\sim$ 0.1 mag and $V$ between 23.5 to 25.0 mag. Finally, \citet{2012ApJ...744...96O} considered the existence of at least two populations with different metallicity and spatial distribution, although they did not find any differences in terms of the age of these populations. 

In this paper, we present a new study of CVnI and its stellar population supported by very deep observations taken with the Advanced Camera for Surveys (ACS) of the Hubble Space Telescope (HST), reaching well below the oldest main sequence turn-off (oMSTO) in the CMD, together with the use of state-of-the-art CMD fitting tools and techniques. This paper is organised as follows. Section \ref{sec:data_red} presents the data used and describes the reduction and photometry techniques; Section \ref{sec:CVnI_CMD} describes the observed CMD and analyses its main characteristics with the help of the superposition of theoretical isochrones. Section \ref{sec:CVnI_SFH} describes the method used for the calculation of the SFH and the results obtained for CVnI. Finally, Section \ref{sec:Discussion} discusses the results and Section \ref{sec:Summary} presents the conclusions and the final summary. Details of the tests performed to estimate the robustness of the results and their temporal resolution, along with additional figures, are described in the appendix.

\begin{table}
	\centering
	\caption{Compilation of the main CVnI parameters}
	\label{tab:char_table}
 \scalebox{0.9}{\begin{tabular}{lccr} 
		\hline
		Parameter & & Value & \\
		\hline
		RA & $\alpha$$_{0}$ & 13h 28m 01.4s  & (5)\\
		Dec & $\delta$$_{0}$ & 33$^{\circ}$ 33$^{'}$ 07.7$^{''}$   & (5)\\
	    Distance & D & 216$\pm$18 kpc   & (4)\\
        Ellipticity & $\epsilon_{exp}$ & 0.45 $\pm$ 0.02 & (1)\\
        Position angle & PA & 80$^{\circ}$ $\pm$ 2$^{\circ}$  & (1)\\
        V-band Absolute Mag. & M$_V$ & $-$8.8 $\pm$ 0.06 mag & (1)\\
        V-band luminosity & L$_V$ & 2.3 $\pm$ 0.3 $\cdot$10$^{5}$ L$_{\odot}$ & (6)\\
        Cent. surf. bright & $\mu$$_{0}$ & 27.1 $\pm$ 0.2 mag arcsec$^{-2}$  & (6)\\
        Half-light radius & $r_{h,exp}$ & 474 $\pm$ 13 pc & (1)\\
        Heliocentric velocity & $v_{los}$ & 30.9 $\pm$ 0.6 km/s & (3)\\
        Velocity dispersion & $\sigma$$_{los}$ & 7.6 $\pm$ 0.4 km/s & (3)\\
        Total Mass & M & (2.7$\pm$0.4)$\cdot$10$^{7}$ M$_{\odot}$ & (3)\\
        Average metallicity & [Fe/H] & -1.98 $\pm$ 0.01 dex  & (2)\\
    
		\hline   
  
	\end{tabular}}
\tablebib{(1) \citet{2018ApJ...860...66M};  (2) \citet{2012AJ....144....4M}; (3) \citet{2007ApJ...670..313S}; (4); Pérez-García (in prep);  (5) \citet{2012ApJ...744...96O}; (6) \citet{2008ApJ...672L..13M}
}
 
\end{table}

\begin{table*} 
\centering
\caption{HST/ACS CVnI\/ observations}
\label{tab:obs}
\begin{tabular}{llllllllll}
 Dataset & Field & RA & Dec & Visit    & Orbits & Instrument & Filter & Exp. Time &  Date \\
         &       & (J2000) & (J2000)  & &       &            &        &    (s)    &       \\
\hline
 JDT301010 & outer & 13 28 18.294 & +33 35 01.34   & 01 & 2   & ACS/WFC & F606W & 5109 & 2019-05-26 \\
 JDT302010 & outer & 13 28 18.294 & +33 35 01.34   & 02 & 2   & ACS/WFC & F814W & 5109 & 2019-05-26 \\
 JDT303010 & inner & 13 28 02.060 & +33 33 30.88   & 03 & 2   & ACS/WFC & F606W & 5109 & 2018-12-29 \\
 JDT304010 & inner & 13 28 02.060 & +33 33 30.88   & 04 & 2   & ACS/WFC & F814W & 5109 & 2018-12-29 \\
\hline
\end{tabular}
\end{table*}

\begin{figure}
\raggedright
\includegraphics[width= \columnwidth]
{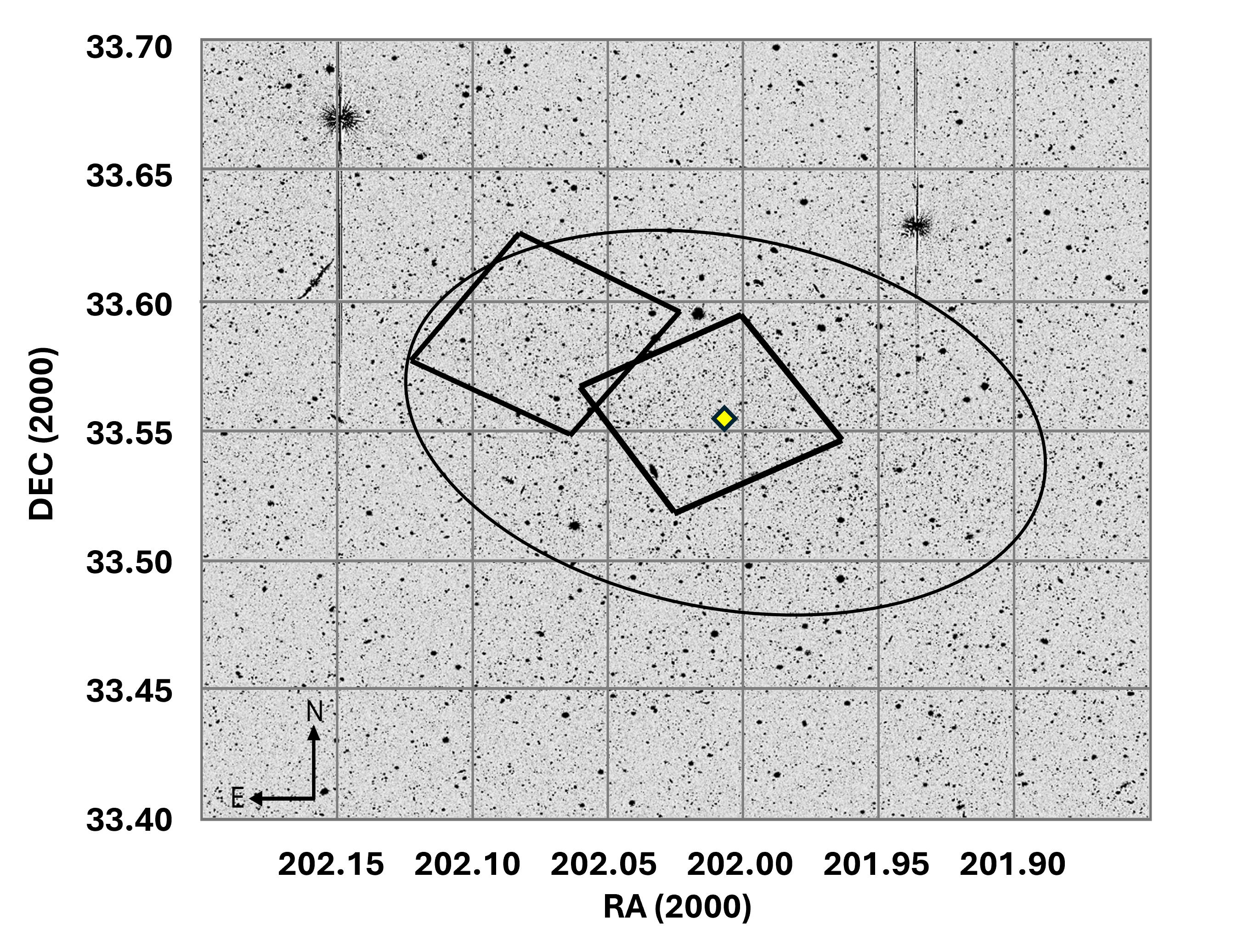}   
\caption{Location of CVnI in RA and Dec, with HST/ACS  targets indicated by the black boxes. The ellipse has a 432 arcsec radius reaching the $r_h$ of CVnI and has been drawn using the ellipticity, PA, and the centre of the galaxy (yellow diamond) indicated in Table \ref{tab:char_table}. The background image is the central section of a mosaic made with the 29 best of 1388 images taken from P.B. Stetson's homogeneous photometry collection.} 
\label{fig:Photometry_areas} 
\end{figure}

\section{Data and data reduction}\label{sec:data_red}

\subsection{Data}\label{sec:data}

This work is based on archival HST/ACS data from the GO program 15476\footnote{Cycle 25, P.I. D. Weisz}. The observations were designed to obtain deep photometry, with stars at the oMSTO measured with S/N of at least 10. Two fields were observed: one close to the centre of CVnI (RA = 13h28m18.3s, DEC = 33$^{\circ}$35$^{'}$1.3$^{''}$, hereafter inner field), and the second contiguous toward the East along the semimajor axis (13h28m2.1s, 33$^{\circ}$33$^{'}$30.88$^{''}$ (hereafter, referred to as the outer field), but still within the half-light radius of CVnI. The configuration is shown in Fig. \ref{fig:Photometry_areas}, while the observations are described in Table \ref{tab:obs}. Two visits of two orbits each were allocated to each pointing, for a total of eight orbits. Each visit did observations in one filter only, with the first visit in F606W and the second visit in F814W. Short and long exposures were collected in both filters (60 sec and $\sim$1100-1300 sec per filter, respectively) to minimise potential saturation issues of bright stars. The total exposure time was 5109 s in both filters of each field. The observations of the inner field were executed in December 2018, while the outer one was observed in May 2019.

\subsection{Data reduction and calibration}\label{sec:red}

Following the steps described in \citet{2010ApJ...722.1864M}, the photometry and the artificial star tests were performed with the DAOPHOT/ALLFRAME suite of codes \citep{1987PASP...99..191S,1994PASP..106..250S}, reducing the $\_flc$ images of the two fields separately. These images are bias-corrected, dark-subtracted, flat-fielded and corrected for the charge transfer efficiency. In brief, the procedure first performs photometry on individual images, modelling the PSF by carefully selecting  bright stars spread all over the field of view in order to take into account spatial variations. Subsequently, a stacked median image is generated and used to build the list of stars that is given as input to ALLFRAME. The process was repeated twice.

The photometric catalogues of the single images were calibrated to the VEGAMAG system and then averaged to obtain a unique list. For subsequent analysis, this was cleaned from extended background objects using a cut in the sharpness parameter; namely, |sha|$<$0.1. This resulted in 11,386 bona fide stars (5,826 stars in the inner field and 5,560 stars in the outer field, with only the latter including the few tens of sources in the overlap region).

As a last step, the photometry was calibrated to the Johnson system taking advantage of the ground-based catalogue of CVnI available from Peter Stetson's homogeneous photometry collection. By using local standard stars, 50 for outer field and $\sim$90 for inner field, we followed the procedures described, for example, in \citet{2011PASP..123..481S} and \citet{2005PASP..117.1049S}, to derive the following transformation equations:
\begin{gather}
V−606 = C_0 + C_1(606−814),\\
I-814 = C_{0}^{'} + C_{1}^{'}(606−814).
\end{gather}
In which the determination of the coefficients $C_0$, $C_1$, $C_{0}^{'}$ y $C_{1}^{'}$, was carried out for each field separately by means of a weighted linear regression, with the inverse of the individual variance of the residuals as the weight.

The coefficients obtained were $C_0$ = $-$0.01, $C_1$ = 0.35, $C_{0}^{'}$ = 0.016, and $C_{1}^{'}$ = 0.02 for the outer field and then $C_0$ = 0.011, $C_1$ = 0.30, and $C_{0}^{'}$ = 0.026, $C_{1}^{'}$ = $-$0.033 for the inner field.

\subsection{Artificial star tests}\label{sec:artifical_star_tests}

Tests with artificial stars are crucial for estimating the photometric errors and the completeness of the final catalogue. They are used to simulate the overall observational errors in the synthetic CMD used in the SFH derivation. Artificial star tests consist of injecting mock stars in the images and repeating the photometric process identically as for the real data. To generate the list of artificial stars we created a synthetic CMD similar to the one later used for the SFH derivation (see Section \ref{sec:SFH_deriv}), covering the full expected ranges of ages and metallicity of the observed galaxy. This allowed us to sample the CMD giving more weight to the more populated sequences. We injected of the order of 10$^{6}$ stars per chip, in multiple runs, to obtain a large amount of statistical data. Stars are distributed on a regular grid on the image, in such a way that artificial stars do not overlap and do not alter the original crowding. Each set of images is processed the same way as the originals. The only exception is that the PSF is recalculated on the basis of a smaller number of stars (80-95\% randomly selected from the list of stars originally used for each image). This prevents stars from being injected and recovered with exactly the same PSF, which could under-estimate the errors. The information saved and required to simulate the observational errors in the synthetic CMD used to derive the SFH are the input and output magnitudes of recovered stars, as well as the input magnitude of lost stars (with output magnitude flagged as 99.999). The same |sha|$<$0.1 cut has been applied to the artificial stars. Details on the use of these artificial stars are given in Section \ref{sec:dispar}.

\section{CVnI CMD}\label{sec:CVnI_CMD}

\subsection{CMD overview}\label{sec:CMD_overview}

\begin{figure}
\raggedright
\includegraphics[width= \columnwidth,height=0.9\columnwidth]
{ 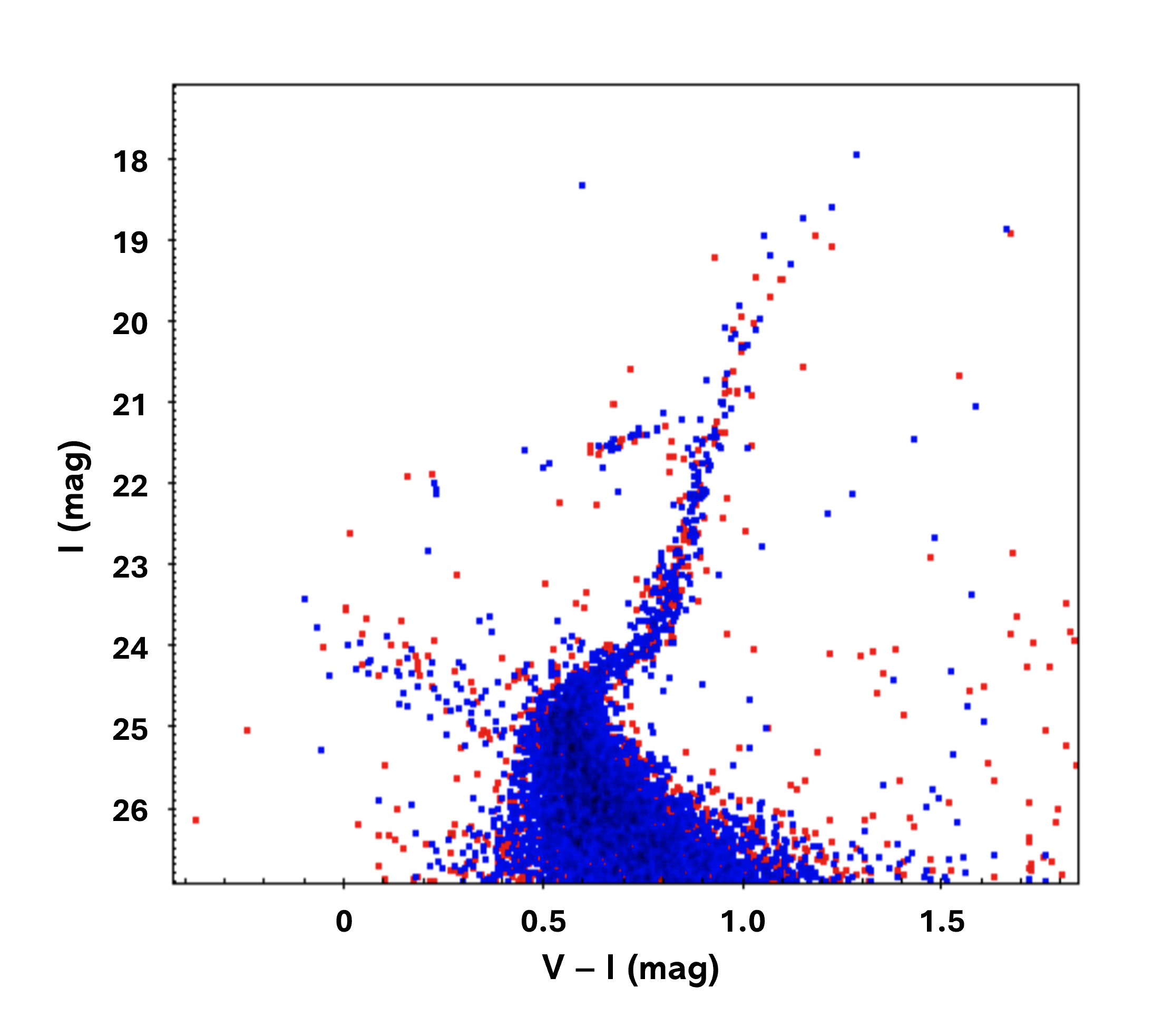}  
\caption{CMD of CVnI obtained from the observations of the two ACS fields 
calibrated to the Johnson system (without yet applying extinction and de-reddening). Blue dots correspond to the inner field and red dots to the outer field.} 
\label{fig:CMD_zoom} 
\end{figure}

Figure \ref{fig:CMD_zoom} shows the CMD of CVnI. Blue and red symbols show sources detected in the inner and outer fields, respectively. The ACS data, being able to resolve very faint stars, allow us to obtain a CMD that spans about nine magnitudes from the tip of the red giant branch (TRGB), I $\sim$ 18 mag, to about 2.5 magnitudes below the oMSTO. 

A visual inspection reveals well-defined oMSTO and subgiant branch (SGB) regions, whose overall shape suggests that CVnI is composed of a mostly old stellar population, with no apparent existence of young stars, as is common in fast-type galaxies \citep{2015ApJ...811L..18G}. Despite a predominantly red horizontal branch (HB), there are no signs of the existence of a red clump, which could have indicated an intermediate-age population. A well-defined and somewhat broad red giant branch (RGB) could be indicative of a possible dispersion in metallicity. An apparent slight broadening between 19 $<$ I $<$ 20.2 mag could be due to the contribution of the asymptotic giant branch.

A faint blue sequence of stars is visible at 25 $<$ I $<$ 23.5 mag and 0 $<$ (V-I) $<$ 0.5 mag. Despite initial suggestions based on shallower photometry of the presence of a young population \citep{2008ApJ...674L..81K, 2008ApJ...672L..13M}, the morphology of this sequence strongly suggests the presence of blue straggler stars (BSS). \citet{monelli12a} relied on the linear correlation between the logarithmic frequency of BSSs normalised to the number of HB stars ($F_{BSS}$ = log($N_{BSS}$/$N_{HB}$)) and the absolute visual magnitude, $M_{V}$, for nearby dwarf galaxies found by \citet{2007A&A...468..973M}, as an independent indicator to identify the blue plume stars of Cetus and Tucana as BSSs rather than young stars. Obtaining the frequency value for CVnI, $F_{BSS}$ = 0.22 $\pm$ 0.08, and with its $M_{V}$ = -8.8 taken from Table \ref{tab:char_table}, when placed in Figure 2 of \citet{2007A&A...468..973M}, we find that it would be reasonably aligned with the correlation found for other galaxies in the LG, reinforcing the assumption that the stars in the blue plume of CVnI are indeed BSSs (even considering the uncertainty in properly defining the HB and BSS selection boxes, which would be reflected in the reported error of $F_{BSS}$, derived as the propagation of the Poisson errors on the star counts). This assumption is fully in line with what is found in all old nearby dwarf spheroidal (dSph) galaxies \citep{mapelli07, mapelli09, monelli12a}. 

\subsection{Isochrone analysis}\label{sec:isochrone_analysis}

\begin{figure}
\raggedright
\includegraphics[width= \columnwidth,height=0.9\columnwidth]
{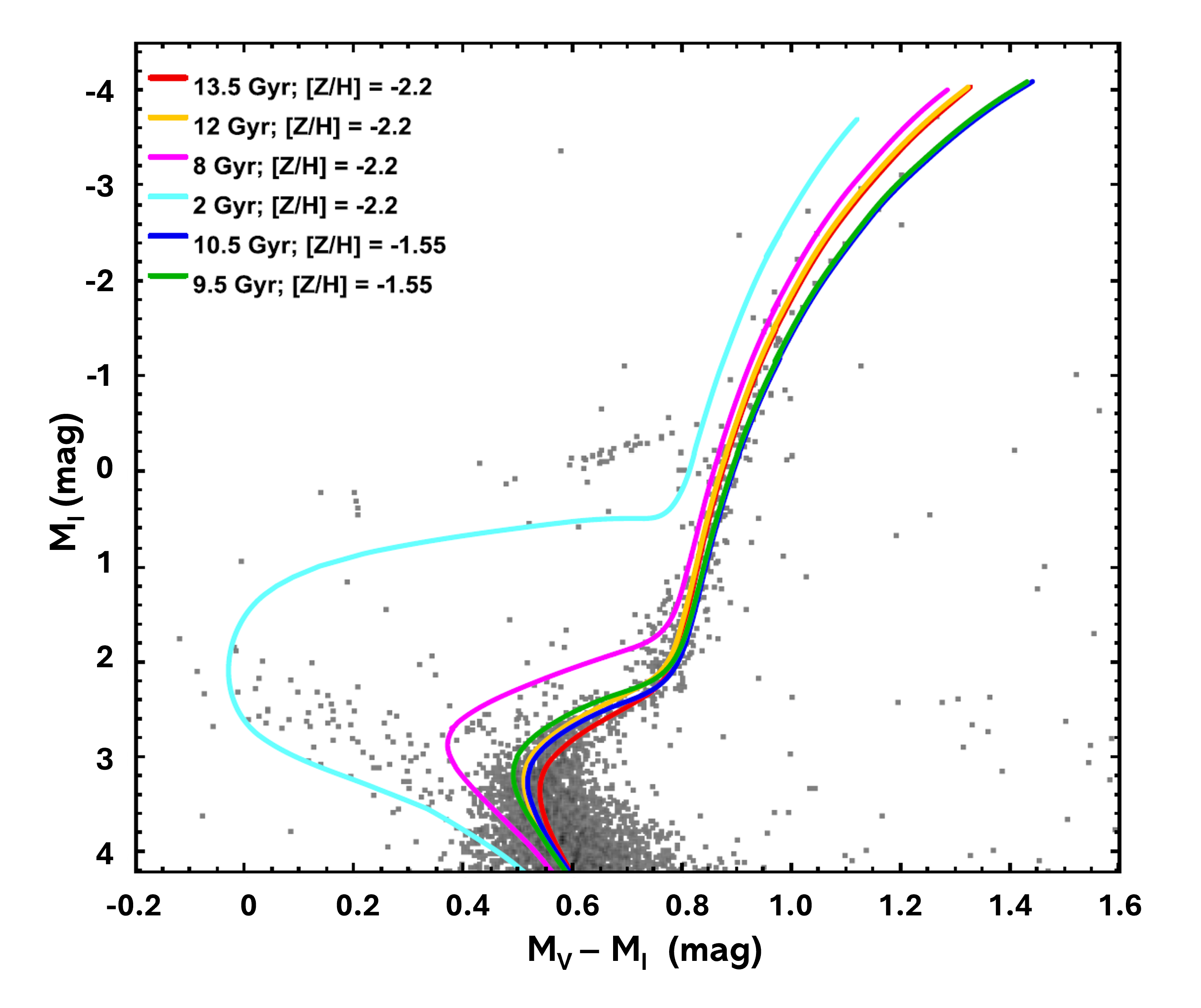}  
\caption{CMD of CVnI superimposed with a selected sample of isochrones for the labelled initial stellar ages and global metallicity. See text for details.} 
\label{fig:isoc} 
\end{figure}

A more detailed description of the stellar populations in CVnI can be derived superposing theoretical isochrones. This is shown in Fig. \ref{fig:isoc} which presents the comparison with BaSTI-IAC\footnote{http://basti-iac.oa-abruzzo.inaf.it/}  solar-scaled isochrones \citep{2018ApJ...856..125H},
with [He] = 0.247, overshooting and diffusion. We considered isochrones at the edges of the range of metallicities found in the literature for CVnI, in both spectroscopic and photometric studies: Z = 0.0001 ([Z/H] = $-$2.2)\footnote{We use [Z/H] to denote global metallicity compared to that of the Sun} and Z = 0.000443 ([Z/H] = $-$1.55) to test the possible existence of stars with these extreme metallicities. To shift the observed diagram to the absolute plane, we corrected for dust extinction assuming a reddening value of E(B$-$V) = 0.014 $\pm$ 0.026 mag \citep{2012AJ....144....4M}, and we assumed the distance derived by Pérez-García et al. (in prep), using a sample of 105 RRLyrae stars (82 new discoveries) identified using ground-based photometry from the collection of 
homogeneous photometry by P. B. Stetson \footnote{https://www.canfar.net/storage/vault/list/STETSON/homogeneous/}. Using Period-Wesenheit relations in the Johnson-Cousins B and V bands (PW(V, B$-$V) relations), they calculated a true distance modulus of $\mu_0$ = 21.69 $\pm$ 0.05 mag, in excellent agreement with previous values in the literature \citep[see e.g.][]{2006ApJ...643L.103Z, 2008ApJ...674L..81K, 2008AJ....135.1361D, 2008ApJ...685L..43K,2012ApJ...744...96O}.

We performed a systematic review of all ages of the chosen metallicities by superimposing them on the CMD of the two ACS fields. Figure \ref{fig:isoc} shows that the isochrones that best match the main sequence (MS), oMSTO, SGB, and RGB zones are those of ages older than 9.5 Gyr in both metallicities, supporting the notion that the dominant population is composed of old stars and presents a significant dispersion in terms of metallicity. 

The bluer and brighter edges of the turn-off and SGB regions suggest the possible existence of younger populations. This is supported by the comparison with the 8 Gyr and [Z/H] = $-$2.2 isochrone (pink curve), which could reveal a possible secondary turn-off, as well as by the comparison with an isochrone of 2 Gyr and [Z/H] = $-$2.2 (cyan curve), which fits the bluer edge of the MSTO reasonably well and could suggest the existence of a young and more metal-rich population (up to 1 Gyr). As already mentioned, it is important to note that we have assumed these stars are BSS; however, these potential young populations are specifically analysed in the next section.

Although these results are very illustrative, the comparison with theoretical isochrones only provides qualitative information. A quantitative description of the SFH in terms of mass formed as a function of age and metallicity will be derived with the CMD fitting technique, as detailed in the next section.

\section{The SFH of CVnI}\label{sec:CVnI_SFH}

\subsection{Overview of the technique}\label{sec:sfh_summary}

In this section, we summarise the basic information regarding the CMD-fitting technique, which is widely used to quantitatively derive the SFH of a stellar system \citep{1991AJ....102..951T, 1996ApJ...462..672T, 2002MNRAS.332...91D, 2010AdAst2010E...3C, 2009AJ....138..558A}. We define the SFH as the amount of mass that has been transformed into stars as a function of time (t) and metallicity (Z). 

The methodology followed here closely matches that described in \citet{2024A&A...687A.168G} for the Chronogal project, particularly for the computation of the synthetic CMD and the fitting of the observed CMD. We use ChronoSynth, DisPar and dirSFH codes and the three basic steps of the process performed with them are: i) ChronoSynth generates the synthetic CMD; ii) DisPar \citep{2021MNRAS.501.3962R} simulates the completeness and the photometric errors; and iii) DirSFH performs the CMD fitting to derive the SFH. This is the first time in which dirSFH is applied to calculate the SFH of a dwarf galaxy.

\subsection{Synthetic CMD}\label{sec:cmdsynth}

We generated a synthetic CMD up to $\sim$10$^8$ stars. To do so, we adopted the following parameters in ChronoSynth:
\begin{itemize}
\item A constant SFR at ages between 13.7 and 0.02 Gyr and metallicities between Z = 0.00002 ([Z/H] = $-$2.9 dex) and Z = 0.003 ([Z/H] = $-$0.7 dex) to cover any possible real population;
\item The fraction of unresolved binary stars, $\beta$ = 1, and the minimum mass ratio of binary stars, \textit{$q_{min}$} = 0.4;
\item The IMF defined by \citet{1993MNRAS.262..545K};
\item As for the isochrone analysis, the selected stellar evolution library is the BaSTI-IAC library using the solar-scaled mixture \citep{2018ApJ...856..125H}, including overshooting and diffusion;
\item The Johnson-Cousins photometric system, assuming a limit magnitude equal to $M_I$ = 4.5 mag;
\end{itemize}

\subsection{Observational errors and completeness}\label{sec:dispar}

We used the DisPar code to simulate the observational errors in the synthetic CMD, using the file generated as the result of the artificial star tests previously described (see Section \ref{sec:artifical_star_tests}). The operation of the code is detailed in \citet{2021MNRAS.501.3962R} and here we provide a brief description of our specific application. The working criteria used by the tool are configurable. For the purposes of this work, we set the following: for each star of magnitudes ($M_{Vs}$, $M_{Is}$) in the synthetic CMD, DisPar searches among all stars in the artificial star archive for at least ten stars within a colour-magnitude box [0.04, 0.1] mag and with input and recovered magnitudes ($M_{Vi}$, $M_{Ii}$) and ($M_{Vr}$, $M_{Ir}$), respectively. If the limit of ten stars is not reached, the search starts again with the box size doubled. Among all the stars found in the colour-magnitude box, one is selected at random and the magnitudes of the considered synthetic star are modified by $M_{Vs}$ = $M_{Vs}$ $-$ ($M_{Vi}$ $-$ $M_{Vr}$) and $M_{Is}$ = $M_{Is}$ $-$ ($M_{Ii}$ $-$ $M_{Ir}$). If the selected artificial star has not been recovered, in which case its ($M_{Vr}$, $M_{Ir}$) are flagged (99.999, 99.999), the synthetic star is also flagged in this way and effectively removed from the CMD, thereby simulating the incompleteness in this way.  

\subsection{Derivation of the SFH}\label{sec:SFH_deriv}

To calculate the SFH, we used the new CMD fitting package called DirSFH, which is a state-of-the-art tool developed as an evolution of previous tools such as IAC-pop \citep{2009AJ....138..558A} or TheStorm \citep{2015MNRAS.453L.113B, 2018MNRAS.477.3507B}. A detailed description of this new tool is given in \citet{2024A&A...687A.168G}, but we summarise its main features and novelties below:
\begin{itemize}
\item  DirSFH requires the definition of an array of values in age and metallicity (seed points) to form a grid that is used to segment the mother CMD by means of a semi-random Dirichlet-Voronoi tessellation \citep{10.1093/comjnl/21.2.168}. This array is used to define the size of the simple stellar populations (SSPs), and thus the resolution in age and metallicity over the whole interval of both. From this array, different sets of SSPs are constructed and one SFH is derived from each set by determining the set of weights to each population that results in the best fit to the observed CMD. At the end of the process, the final SFH is obtained as the average of all of these individual best fit SFHs.
\item  It allows the user to define an area in the observed and in the mother CMDs (the same for both, which will be called bundle hereafter) that will delimit the area in both of them that will be used for the fit. Within that area, the process defines a grid of ‘pixels’, typically of a size of 0.2 mag in magnitude and 0.1 mag in colour, and calculates a magnitude-colour histogram with it. This grid is then shifted in 5x5 steps in colour and magnitude to generate new histograms with each of the 25 steps, resulting in a CMD with oversampled pixels. Finally, the process generates a matrix of weights, assigning one to each pixel as the inverse of the variance of the stellar ages in it, so that the more ‘unique’ a pixel is in terms of the age of its stars, the greater its weight.
\item Small systematic offsets in colour and magnitude between the real data and those generated from the theoretical models might exist. To estimate them, DirSFH runs several SFH calculations with the mother CMD shifted in the colour-magnitude space within a given range. Then analyses the resulting residuals after each calculation and chooses the best shift as the weighted average of those resulting in the lowest residuals.
\item DirSFH allows the user to define several runs of the complete SFH calculation to obtain the final result as a weighted average of all of them.
\end{itemize}

To potentially cover the ranges of age and metallicity presumably present in the real population, we chose the following arrays in age and metallicity (seed points) for all SFH calculations performed in this work:
\begin{itemize}
\item Ages (Gyr) = 0.02, 0.05, 0.1, 0.5, 1, 1.5, 2, 2.5, 3, 3.5, 4, 4.5, 5, 5.5, 6, 6.5, 7, 7.5, 8, 8.5, 9, 9.5, 10, 10.5, 11, 11.5, 12, 12.5, 13, and 13.7;
\item Z = 0.00002, 0.00005, 0.0001, 0.0003, 0.0005, 0.0007, 0.001, 0.002, 0.003 (equivalent to [Z/H] = $-$2.9, $-$2.5, $-$2.2, $-$1.7, $-$1.5, $-$1.35, $-$1.2, $-$0.9, and $-$0.7 dex).
\end{itemize}

\begin{figure*}
\sidecaption
 \includegraphics[width=12cm]{ 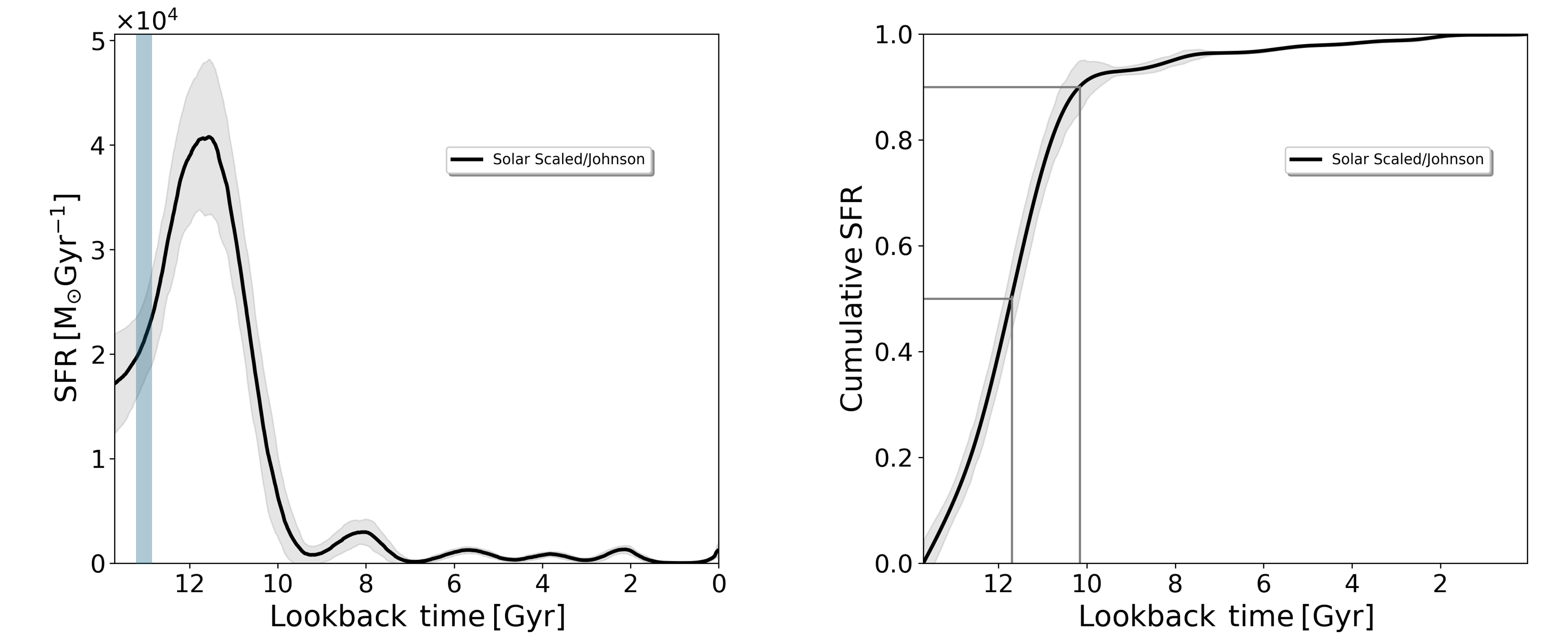} 
  \caption{Left: SFR of CVnI, showing that the galaxy had active star formation up to $\sim$9 Gyr, with the peak at $\sim$11.8 Gyr, followed by a final short period of low intensity at 8 Gyr. The small peaks at younger ages are not considered significant. (see text for full details). The blue shaded area marks the approximate duration of the reionisation epoch \citep[6 $\leq$ $z$ $\leq$ 9,][]{2022ARA&A..60..121R}. Right: Cumulative SFR, showing that CVnI had transformed 90\% of its mass into stars around 10 Gyr.}
  \label{fig:SFR_CVnI_Johnson_1} 
\end{figure*}

\subsection{Results: CVnI SFH}\label{sec:Result_CVnI_SFH}

Figure \ref{fig:SFR_CVnI_Johnson_1} shows (see left panel) the SFR, namely, the amount of mass transformed into stars per unit time. CVnI had the strongest star formation activity at old ages, with the highest peak at around 11.8 Gyr and after which the SFR rapidly declined to quench almost completely $\sim$9 Gyr ago. The main peak is followed by a second, less intense and shorter episode at around 8 Gyr ago. Some minor activity seems to be present at 6, 3, and 2 Gyr. The right panel of the same figure shows the cumulative SFR, where the two grey lines indicate the 50th and 90th percentiles, showing, respectively, that 50\% of the mass transformed into stars before 11.8 Gyr ago, and 90\% before 10 Gyr ago. Thus, we can see that the stellar population of this galaxy consists mainly of old stars. 

\begin{figure*}
\centering 
\includegraphics[width=0.9\textwidth]{ 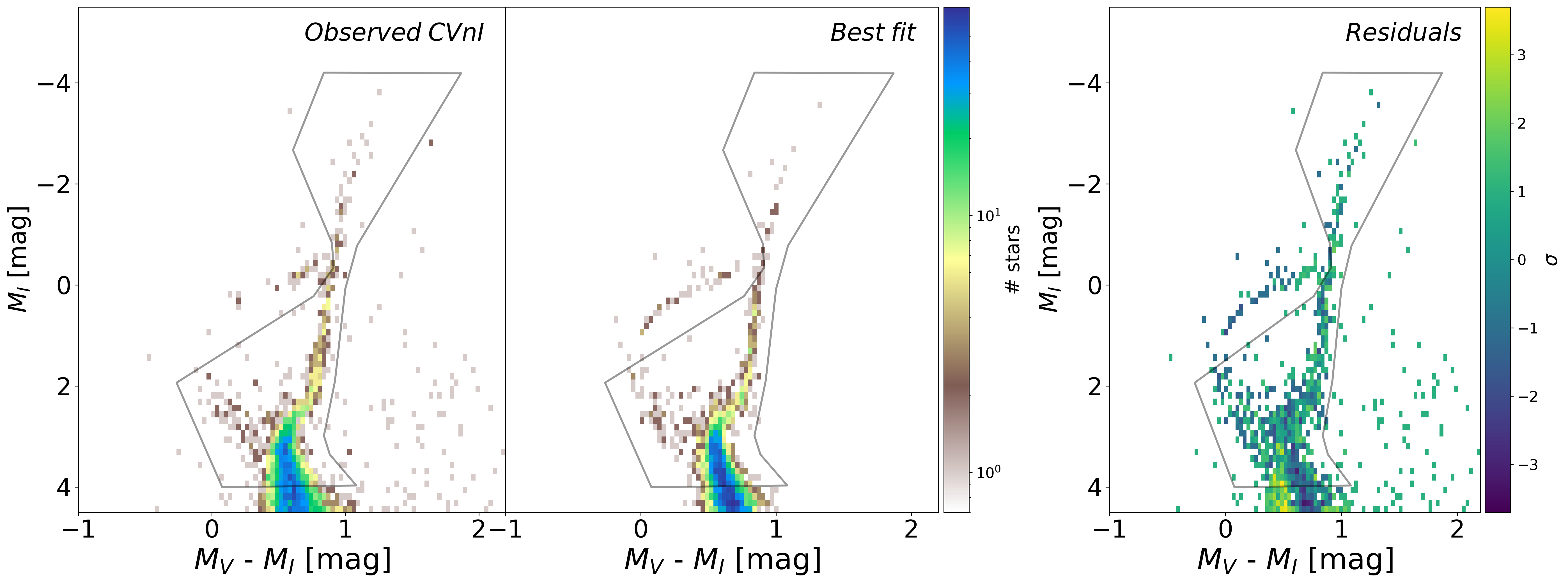} \\
\caption{Solution for CVnI CMD calculated with dirSFH in Johnson photometry, for all the stars in both ACS fields. It had been used the mother CMD described in Section \ref{sec:cmdsynth}, with solar-scaled mixture and the seed points for age and metallicity described in Section \ref{sec:SFH_deriv}.  The shape and size of the bundle were selected to appropriately cover the areas that provide the most accurate information for the CMD-fitting technique.}
\label{fig:CMD_solution} 
\end{figure*}

Figure \ref{fig:CMD_solution} shows the observed and the solution CMDs (left and centre panels, respectively), with the bundle used superimposed. The shape and position of the bundle were chosen to adequately cover the evolutionary phases from which a reliable SFH can be derived, taking into account that the completeness of the observed data is greater than 90$\%$ within the range of magnitudes covered by it. The right panel shows the residuals of the fit, and it can be seen that in general the fit is of good quality, with most of the pixels showing an agreement at the $\pm$1$\sigma$ level. Larger (but still small) discrepancies at the 2-3 sigma level are visible in the faint main sequence, with most of the discrepancy occurring outside of the bundle, close to the magnitude limit of the photometry, where the photometry errors seem to have been somewhat underestimated. It is interesting to note that in the HB, the difference between the observed and the solution CMD is very noticeable, with the latter being more extended and populated towards the blue side. This can be explained by the mass loss efficiency along the RGB adopted when generating the mother CMD. In particular, it is parametrised with a relatively high average value of $\eta$ = 0.3 (lower mass loss values would result in a HB less extended towards the blue). This has been the primary reason why, after numerous tests, the HB has not been included in the bundle used to derive the solutions, so that potential biases in the results can be avoided, especially in terms of metallicity.

\begin{figure*}
\sidecaption 
\includegraphics[width=12cm]{ 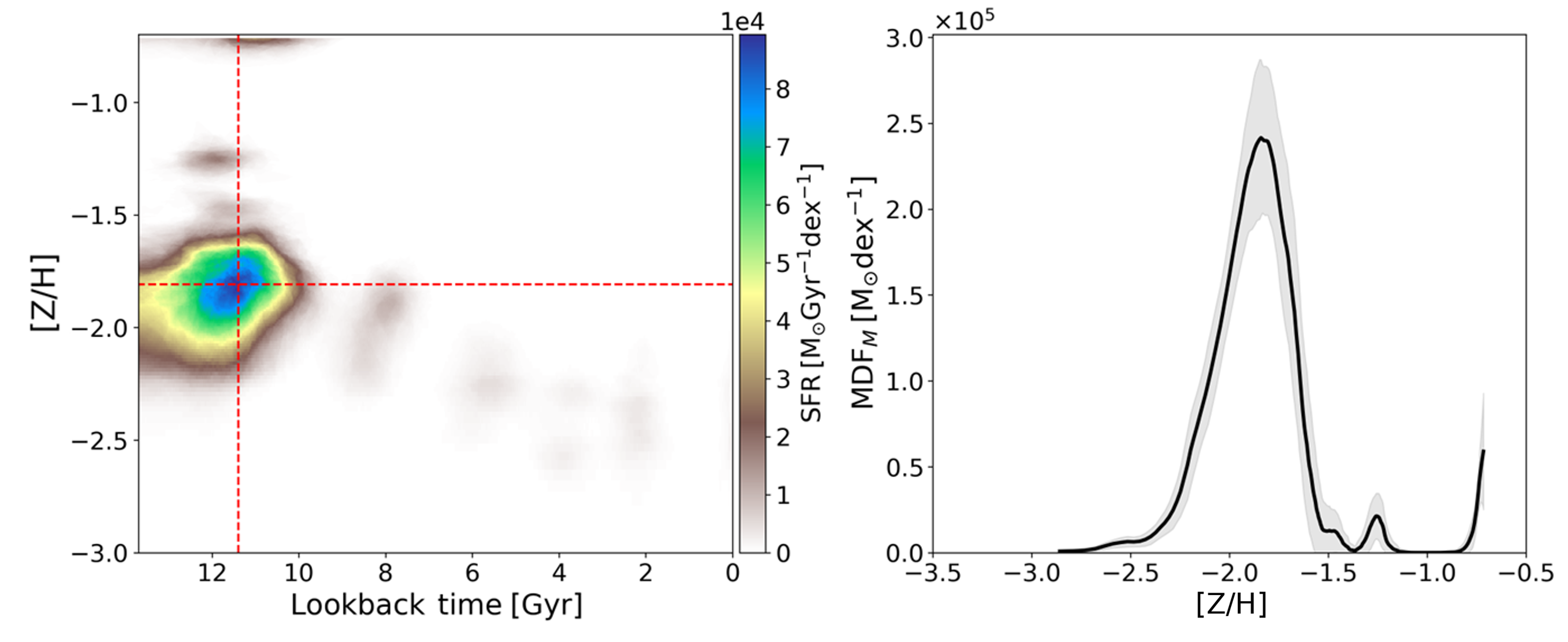}   
    \caption{Left: SFH in the age-metallicity plane shows that CVnI has a unique stellar population with ages mainly greater than 10 Gyr and a metallicity range between $-$2.5 and $-$1.5 dex in [Z/H] (the dashed red lines mark the position of the maximum rate of star formation in age and metallicity). Right: Metallicity distribution function shows clearly that the CVnI stellar population is metal-poor, centred around [Z/H] = -1.8 dex with a dispersion of $\sigma$ $\sim$ 0.16 dex. The small peak at $-$1.3 dex and the tail at high metallicities are considered artefacts due to the presence in the CMD of stars not belonging to the galaxy (see Section \ref{sec:bss}).} 
    \label{fig:AMR_MDF_CVnI_Johnson_1} 
\end{figure*}

\begin{figure*}
\sidecaption
\includegraphics[width=12cm]{ 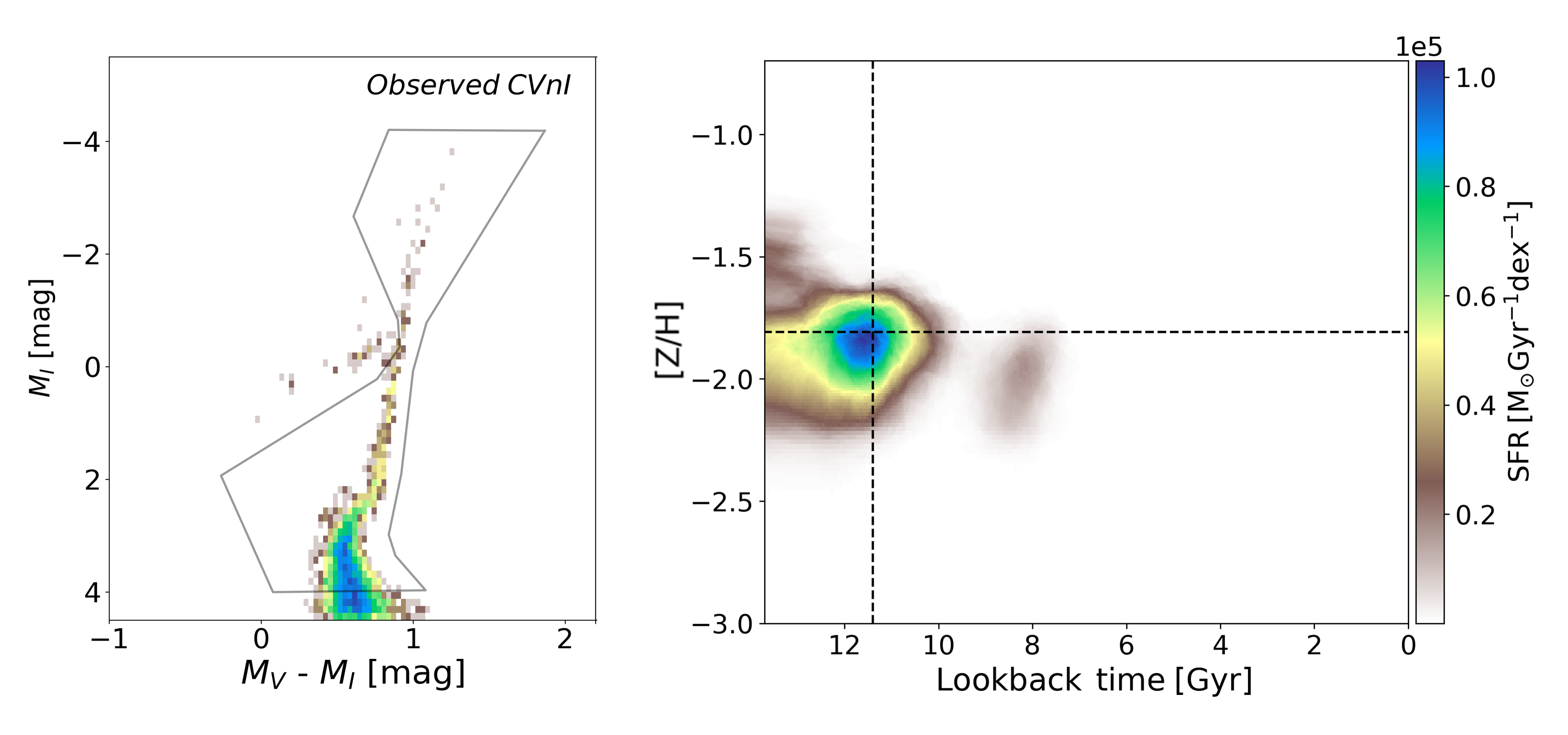}  
    \caption{Left: Observed CMD after removing the stars redder than the RGB, the stars between the HB and the SGB, and the blue plume stars. Right: Age-metallicity plane of the solution obtained with this 'cleaned' population, in which the shadow at old age and high metallicity and the shadows at ages younger than 8 Gyr have disappeared. The dashed lines mark the position of the peak of star formation in age and metallicity showed in Figure \ref{fig:AMR_MDF_CVnI_Johnson_1} for the whole original observed population (see Section \ref{sec:bss} for more details).} 
    \label{fig:Blue_plume_solution} 
\end{figure*}

\begin{figure*}
\centering 
\includegraphics[width=0.9\textwidth]{ 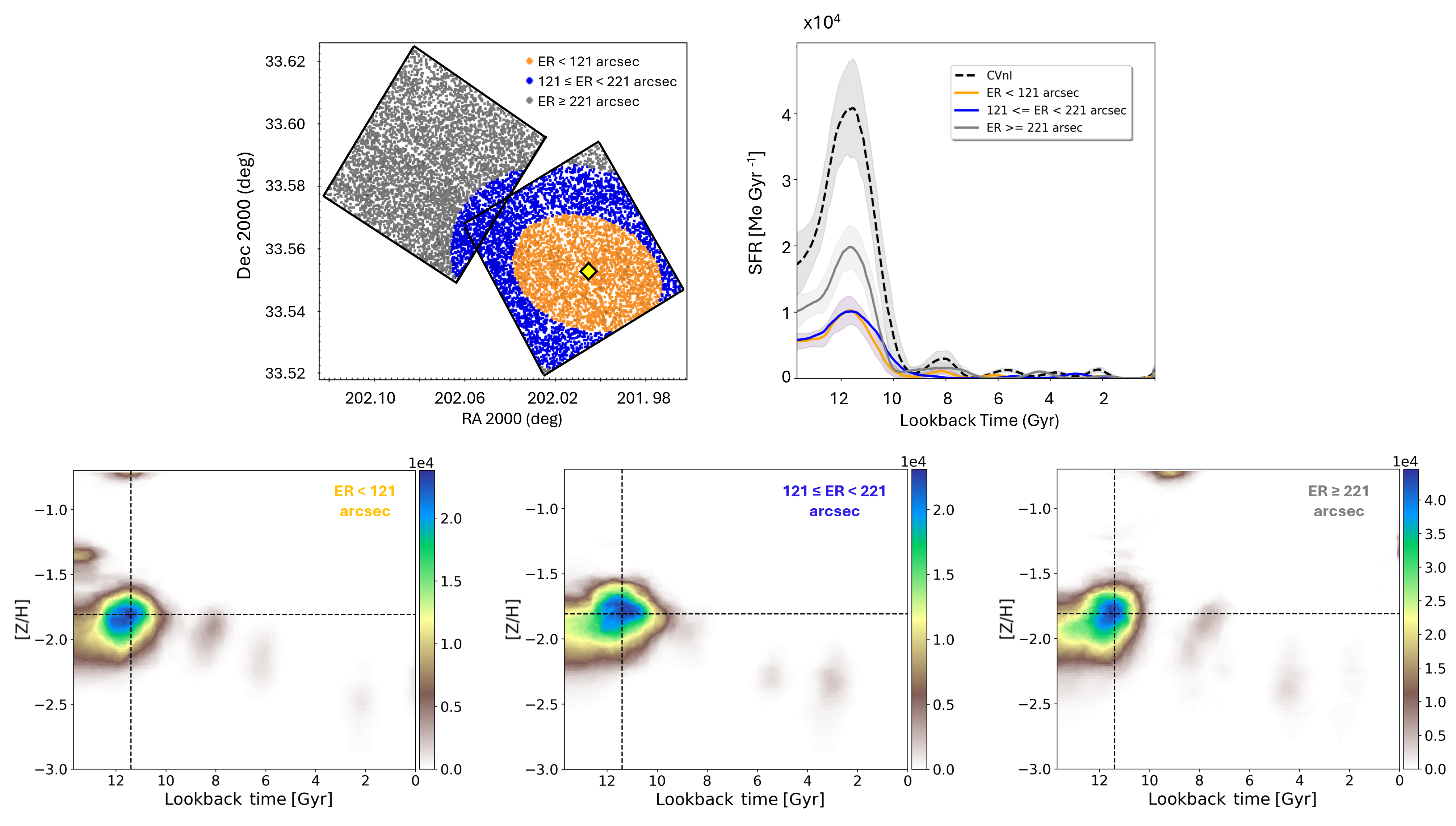} \\   
\caption{Top-left: Division of ACS observed stars into three zones by elliptical radii along the major axis of the galaxy. Top-right: SFR obtained in each of the three zones compared with the SFR obtained for the total of stars. Lower plots: age-metallicity distributions obtained in each zone. In the three plots, the dashed lines mark the position of the peak of star formation obtained for the whole original solution, showing that in all three zones the peaks of star formation are located virtually in the same position. }  
\label{fig:CVnI_AMR_elliptical_zones} 
\end{figure*}

Figure \ref{fig:AMR_MDF_CVnI_Johnson_1} shows the SFH in the age-metallicity plane, which allows us to see the amount of mass that has been transformed into stars by age and metallicity). It can be seen that the population of CVnI is mainly composed of old and metal-poor stars. The dashed red lines mark the star formation maximum in this age and metallicity plane, indicating that the population has its peak at about [M / H] = $-$ 1.8 dex and $\sim$ 11.4 Gyr ago. It can also be seen that the stars corresponding to the 8 Gyr formation period have a mean metallicity very similar to that of the older population. It is interesting to note that the age-metallicity relation (AMR) plot also shows possible populations at ages younger than 8 Gyr, corresponding to those already shown in the SFR graph, as well as other potential high-metallicity populations with an age between 11 and 12 Gyr. The right-hand panel of Figure \ref{fig:AMR_MDF_CVnI_Johnson_1} shows the metallicity distribution function (MDF), which shows the mass that has been transformed into stars per unit of metallicity. We can appreciate that the CVnI population has a relatively broad metallicity width, with a dispersion of $\sigma$ $\sim$ 0.16 dex. This plot also shows a very small peak in [Z/H] around $-$1.3 dex, and another potential high-metallicity peak at [Z/H] > $-$1.0 dex, corresponding to both the old and high-metallicity populations shown in the age-metallicity plot.

\subsection{On the reality of high-metallicity old stars and low-metallicity young stars}\label{sec:bss}

Given the overall AMR described in the previous section, the existence of old metal-rich or young metal-poor stars, which would have required an unusual old infall of very metal-rich gas (or a prompt metal enrichment) as well as a recent infall of metal-poor gas, would be unexpected. To investigate these possible populations in depth, we performed three ad hoc tests, subtracting part of the observed CMD stars from each of them, using the same mother synthetic CMD, the same bundle, and the same parameters as in the original SFH calculation. First, we subtracted all the stars redder than the RGB (136 stars subtracted). Second, from this 'cleaned' CMD, we subtracted the few stars located below the HB and over the SGB, which might reasonably be field stars or galaxies (19 stars subtracted). Finally, to this 'over-cleaned' CMD, we subtracted the blue plume area (71 stars subtracted). The result of this last test is shown in Figure \ref{fig:Blue_plume_solution} (and the results for all three tests are shown in Figures \ref{fig:CMD_NoRGB_NosubHB_noBP} and \ref{fig:SFH_NoRGB_NosubHB_noBP} of the appendix \ref{sec:extra_figures}). It can be seen that the subtraction of the stars has not affected the overall SFR and metallicity characteristics of the galaxy, but has led to the disappearance of traces of the old metal-rich and young metal-poor populations. 

If these populations were really belonging to CVnI, and not foreground stars or background galaxies, they would also be expected to have counterparts in other areas of the CMD, so that the subtraction of these small numbers of stars ($\sim$ 6\% of the total) should not have caused the complete disappearance from the suspicious features in the age-metallicity plane. Thus, these tests reasonably rule out the existence of an ancient population rich in metals and also further support our hypothesis that the blue plume stars are true BSs.

\subsection{Spatial gradients}\label{sec:spatial}

As mentioned in the introduction, previous studies of CVnI have proposed the existence of multiple populations in age  \citep{2006MNRAS.373L..70I, 2008ApJ...674L..81K, 2008ApJ...672L..13M}, and in metallicity, with a possible spatial gradient  \citet{2012ApJ...744...96O}. To investigate whether there might be some kind of gradient or distinct spatial distribution of the stellar population in the galaxy, we computed SFHs as a function of the distance from the centre. Taking advantage of the fact that the ACS observing fields are oriented almost along the major axis of the galaxy, we decided to divide all observed stars into three zones using elliptical radii. The radii of each zone were chosen so that the bundle contained approximately the same number of stars in each zone and the SFH calculation was performed in the same way as for the global solution. To properly reproduce the observational errors, the artificial star files were also divided by the same three elliptical-radius zones, so that after applying the DisPar process, a specific mother CMD has been generated for each zone. 

The SFH solutions obtained for each of these three zones are shown in Figure \ref{fig:CVnI_AMR_elliptical_zones}, where both the SFR and the age-metallicity distribution plots are depicted. It can be seen that in all three zones the period of star formation activity coincides within the error, and the global distribution of stars in the age-metallicity plane remains practically equal. The most notable difference is the disappearance of the 8 Gyr population in the intermediate radius zone, but as it is present in the inner and outer zones, it seems reasonable to consider the existence of a null spatial gradient in age or metallicity, at least on the surveyed area. On the other hand, the metal-poor young stars appear in different positions in each zone, which could indicate that they do not correspond to real stars, reinforcing the hypothesis we argue above. We do note, however, that these results were indeed obtained with observations of very localised areas of the galaxy, close to its centre and along the direction of its major axis, so they cannot be considered conclusive or representative of the entire population of the galaxy. Currently, there are not enough available stars from other areas of the galaxy to carry out a similar CMD-fitting study that might reveal the existence of possible population gradients.

In addition to the tests described above, we also performed a series of robustness tests on the calculation process. The results of these tests (summarised in Table \ref{tab:SFR_mocks} and detailed in the appendix \ref{sec:Robustness}) show that the process is stable to changes in the stellar libraries used, to changes in critical parameters such as the binary fraction, and that its accuracy is very high for determining the ages and metallicity ranges of mock and real populations, such as cluster NGC 6341 (M92). These results serve to guarantee the quality of the solutions and strengthen their reliability.  

\begin{table*} 
\centering
\caption{CVnI\ and\ Mocks\ SFR\ and\ MDF\ characteristics\ (all ages in Gyr)}
\label{tab:SFR_mocks}
\begin{threeparttable}
\begin{tabular}{lcccccccc}
  Object    &	Age\_i  &	Width\_i  &	Mean Age\_o & ${\sigma}$\_o &	[Z/H]$_{i}$ &	${\sigma}$$\_{i}$   &	[Z/H]$\_{o}$ & ${\sigma}$$\_{o}$ \\
\hline
\ CVnI	                & $-$    & $-$     & 11.8   & 0.9  	                & $-$    & $-$     & $-$1.86   & 0.16  \\		
\ NGC 6341 (M92)	                & 13.58 (1)    & --     & 12.9   & 1  	                & -1.77 (1)    & --     & $-$1.9   & 0.15  \\\ Mock 13Gyr (Wide)	& 13	& 1	 & 13.31   & 0.86	& $-$1.83	 & 0.38	  & $-$1.83     & 0.21   \\		
\ Mock 13Gyr (Narrow)	& 13	& 0.001	 & 12.80 	 & 0.65	 & $-$    & $-$ & $-$    & $-$ \\	
\ Mock 12Gyr (Wide)	& 12	& 1	 & 11.85   & 0.55  & $-$1.70	 & 0.32	  & -1.73   & 0.20  \\	
\ Mock 12Gyr (Narrow)	& 12	& 0.001	 & 11.98 & 0.57  & $-$    & $-$ & $-$    & $-$ \\		
\ Mock 11Gyr (Wide)	& 11	& 1	 & 10.97 & 0.5
& $-$1.49	 & 0.22	  & $-$1.48   & 0.13  \\	
\ Mock 11Gyr (Narrow)	& 11	& 0.001	 & 10.96 & 0.46  & $-$    & $-$ & $-$    & $-$ \\		
\ Mock 10Gyr (Wide)	& 10	& 1	 & 10.11 & 0.47  & $-$1.34	 & 0.16	  & $-$1.39   & 0.14  \\
\ Mock 10Gyr (Narrow)	& 10	& 0.001	 & 9.94 & 0.44  & $-$    & $-$ & $-$    & $-$ \\		
\hline \\
\end{tabular} 
\tablebib{ (1) \citet{2025A&A...704A.255A}
}
\end{threeparttable}
\end{table*}

\section{Discussion.}\label{sec:Discussion}

\subsection{Obtained SFH versus the literature.}\label{sec:SFH_vs_lit}

\begin{figure}
\raggedright
\includegraphics[width= 0.9\columnwidth]{ 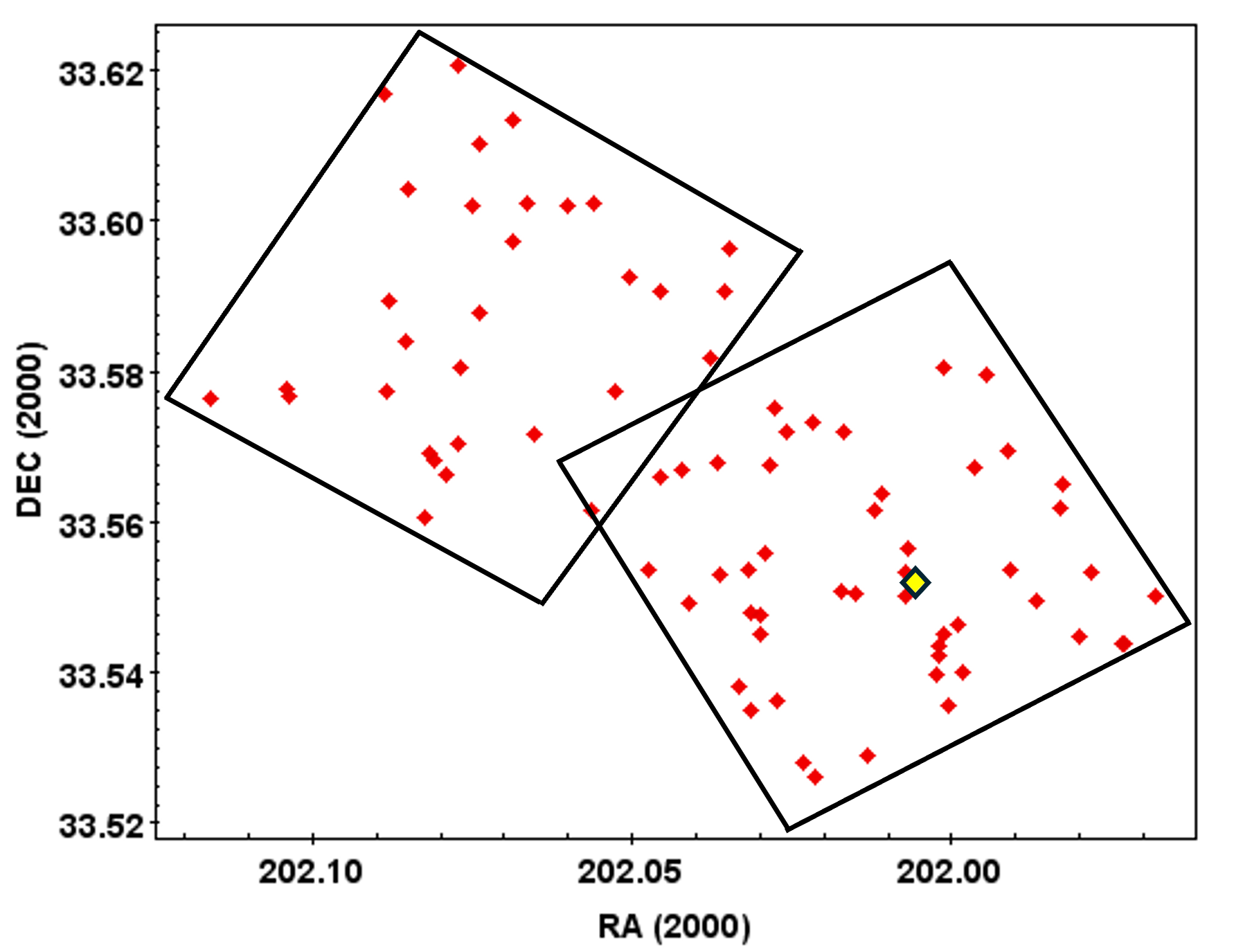}   
\caption{Spatial distribution of the stars located in the blue plume of the CMD (red diamonds). It can be seen that they are smoothly distributed throughout the observed fields (black squares). The yellow diamond marks the centre of CVnI.} \label{fig:BP_stars_dist} 
\end{figure}

The accuracy of the results of the SFH of CVnI reported in Section \ref{sec:Result_CVnI_SFH}, in terms of age and metallicity, supported by the tests carried out with mock populations and M92 (described in the appendices \ref{sec:Mock_pops} and \ref{sec:M92}), indicate that the galaxy had a very active star formation phase at old ages. A main burst of star formation was revealed, with a peak of intensity around 11.8 Gyr ago and whose temporal duration can be estimated at a maximum of $\sim$1.7 Gyr, based on the comparison to the age resolution tests. Therefore, CVnI is dominated by an old stellar population, older than 11 Gyr (with perhaps the exception of the small population of 8 Gyr), in line with the studies published by \citet{2007ApJ...670..313S} or \citet{2012ApJ...744...96O}. As mentioned in the introduction, other authors have inferred the possible existence of populations of stars younger than 8 Gyr. This has been suggested in the results of spectroscopic studies \citep{2006MNRAS.373L..70I,2007MNRAS.380..281M} in which two kinematically distinct stellar subpopulations were identified, associating the more metal-rich and colder one with potential young stars. Other works based on photometric CMD analysis \citep{2008AJ....135.1361D,2008ApJ...672L..13M} have reached similar conclusions. \citet{2008ApJ...672L..13M} studied separately the stars found in the BP of their CMD, finding that they were spatially concentrated within $\simeq$ 2$\arcmin$ to the east and near the centre of the galaxy, also showing some spatial coincidence with the cold component that they had identified in  \citet{2007MNRAS.380..281M}, which led them to estimate that this was potentially a population of young stars. However, the spatial distribution of the stars in the BP of our CMD shows no signs of being spatially concentrated near the centre or in a specific area of the galaxy (see Figure \ref{fig:BP_stars_dist}). On the contrary, they are smoothly distributed throughout the observed fields, supporting our hypothesis that these are not young stars and, rather, BSs instead. Moreover, our study, carried out with the deepest CMD and with the largest number of stars studied to date, has not found any evidence of the possible existence of subpopulations younger than 8 Gyr.

We conclude that the main population of CVnI is mainly metal-poor, although with a significant dispersion. The main peak can be approximated by a Gaussian centred on [Z/H] $\sim$ $-$1.8 and with a dispersion of $\sigma$ $\sim$ 0.16 dex, with extreme values at [Z/H] $\sim$ $-$2.4 and [Z/H] $\sim$ $-$1.5 dex. As with the age results, our metallicity results are in line with those of the studies mentioned in the previous paragraph, with the exception that our study does not show any evidence of populations with a different metallicity range within the area studied in this paper.

\begin{figure}
\raggedright
\includegraphics[width= 0.9\columnwidth]{ 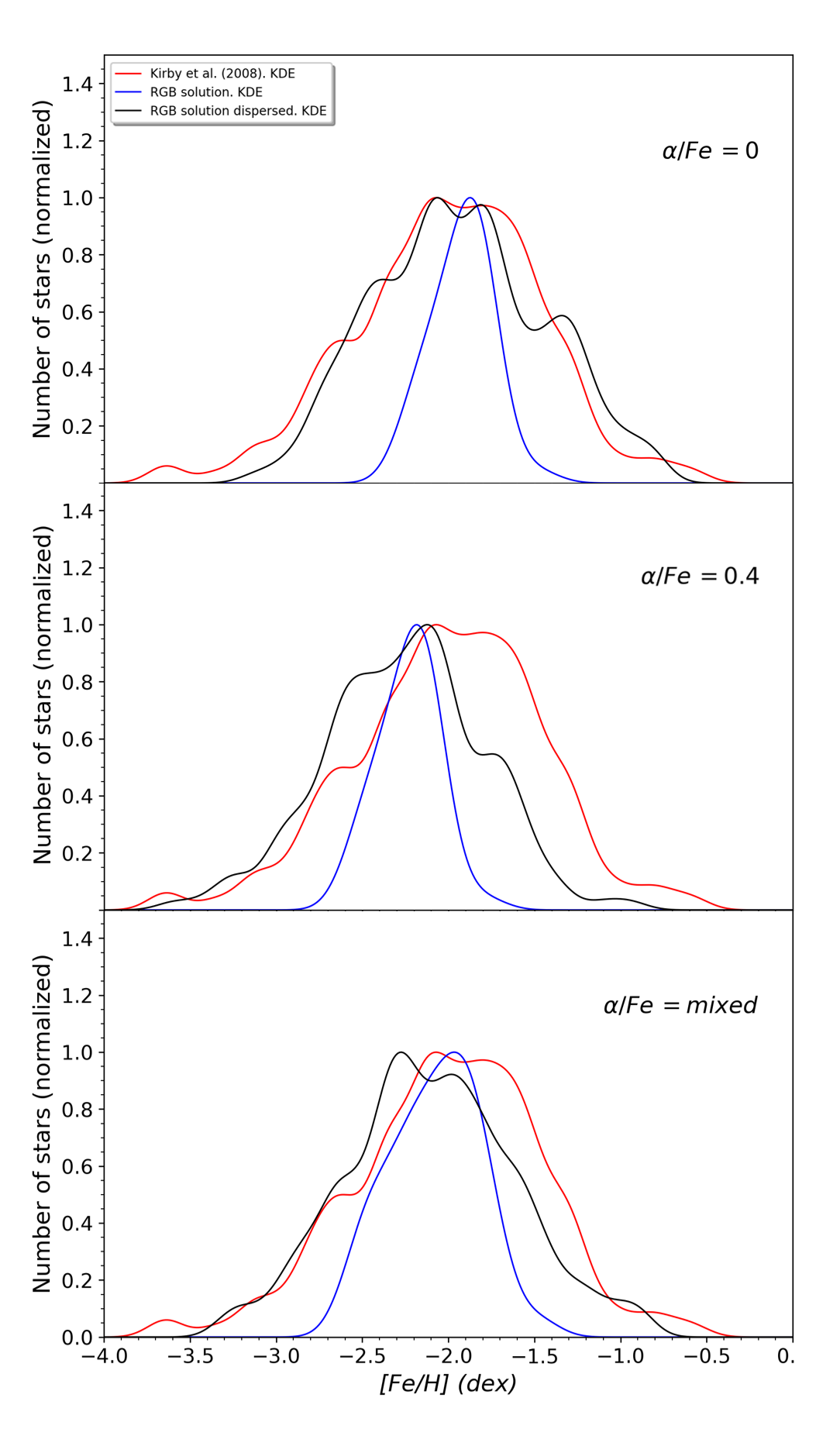}   
\caption{Metallicity of the RGB stars of our solution (blue lines), transformed to [Fe/H] using three possible scenarios of $\alpha$ elements abundance, superimposed on spectroscopic metallicity obtained by \citet{2008ApJ...685L..43K}, (red lines). The black lines represent our solution convoluted with the standard deviation of the spectroscopic distribution.  See text for full details.} \label{fig:Met_Comp} 
\end{figure}

We compared the metallicity of our solution with previously published spectroscopic results. In this regard, the most comprehensive work available is that of \citet{2008ApJ...685L..43K}, which consists of a
spectroscopic study of 165 stars in the galaxy's RGB. To make this comparison consistent, we selected stars from our solution that also belong to the RGB. Since our metallicity is obtained in terms of [Z/H], it was necessary to transform it to iron to make the comparison. Thus, we used the relationship obtained by \citet{2021ApJ...908..102P} as an update of formula (3) of \citet{1993ApJ...414..580S} and expressed as
\begin{gather}
    \hspace{1cm} [\text{Z/H}] = [\text{Fe/H}] + \log(0.694 \cdot 10^{[\alpha/\text{Fe}]} + 0.306)
    \label{eq:metalicidad}.
\end{gather}
\par
\vspace{1mm}
The limitation of this approach is the need to adopt a value, or values, for the abundance of $\alpha$-elements, which in the case of CVnI are not available. Therefore, and taking into account the extreme values and the width in metallicity of the MDF obtained, which indicate a possible complex population in $\alpha$ elements, we devised three possible scenarios for our selection of stars: a pure $\alpha$-enhanced population ([$\alpha$/Fe] = 0.4), a pure solar-scaled population ([$\alpha$/Fe] = 0), and a mixed population, where the abundance of $\alpha$ elements decreases progressively between the above values from the most metal-poor to the most metal-rich stars. In Figure \ref{fig:Met_Comp}, the red and blue curves in each of the three panels represent the smoothed histograms (using KDE with a fixed $\sigma$=0.1) of the spectroscopic [Fe/H] and our solution for each of the scenarios described (red and blue lines, respectively). It can be seen that the mixed scenario agrees best with the spectroscopic data, suggesting that the galaxy experienced a rapid increase in its Fe abundance during its star-forming period. However, there is a difference in the width of the distributions that might be due to the different dispersion of the spectroscopic measurements compared with the precision with which we infer metallicity with the CMD-fitting technique \citep[as discussed and quantified in][]{2024A&A...687A.168G}. The average standard deviation of the spectroscopic metallicities of CVnI is 0.46 \citep{2008ApJ...685L..43K}, while the precision of our MDF can be estimated from the $\sigma$ of the metallicities inferred for M92 ($\sigma_o$ = 0.15 in Table\ref{tab:SFR_mocks}) . We simulated the dispersion of the spectroscopic metallicities in our MDF following \citet{2024A&A...687A.168G}: we added a normal noise distribution to the metallicities of the stars in the solution CMD, with a standard deviation equal to the square root of the quadratic difference of the spectroscopic sigma minus our estimated sigma. The black curves in Figure \ref{fig:Met_Comp} show this new dispersed metallicity of our solution. It is worth noting that the new widths are comparable to those of the spectroscopic distribution, thereby reflecting the reliability of the metallicity distributions inferred with the CMD-fitting technique.

\subsection{Evolutionary scenarios}\label{sec:evol_scenarios}

The SFH of CVnI indicates that the galaxy reached its peak SFR 11.8 Gyr ago and almost completely ceased forming stars approximately 10.9 Gyr ago; notably, taking into account the time resolution of the SFR, found with the mock tests described in Appendix \ref{sec:Mock_pops}. Thus, we want to find out the cause behind this SFH pattern could be. Currently there are three main mechanisms considered, which would be potentially modulating the SFH in small galactic halos: heating by the cosmic UV background, internal feedback due to supernovae (e.g. \citet{1999ApJ...513..142M, 2004ApJ...609..482K, 2010MNRAS.402.1599S, 2017MNRAS.471.3547F, 2020MNRAS.497.1508R, 2020MNRAS.498.4134G}), and possible local interactions with massive objects or in a very dense environment. We examine each of these scenarios in more detail below. 

As already mentioned, CVnI is a distant dSph satellite of the MW, located in a relatively isolated environment. This could have led to it evolving as a slow galaxy, with its star formation being somewhat extended; conversely, its small size (placing it on the borderline between ‘classical’ and UFD) could have led it to be a fast galaxy. Our results reveal that it is indeed the latter case. Its population is fundamentally old, as is the case of other isolated galaxies, such as Tucana \citep{2010ApJ...722.1864M} or Cetus \citep{2010ApJ...720.1225M}, with which CVnI shares a very similar SFH, despite their differences in size, luminosity, and distance. Similarly, CVnI can be considered a fast galaxy \citep{2015ApJ...811L..18G} due to the age of its star formation quenching and this classification is reinforced according to Figure 14 of \citet{2022Univ....8..191M}, where CVnI is clearly located in the region of fast galaxies due to the frequency of its AC stars (Pérez-Garcia et al. in prep), logS = $-$0.36, as a function of its V-band magnitude.

\citet{2015ApJ...811L..18G} postulated that fast galaxies are candidates for having had their star formation severely affected by the effect of UV background combined with internal feedback, to the point of losing all their interstellar medium (which, however, does not necessarily occur by the end of the reionisation epoch). In this scenario, the SFH of CVnI would have been greatly affected by reionisation. However, our study (and all the previously cited publications) show that this has not been the case. In fact, we note that the peak of star formation occurred $\sim$0.8 Gyr after the end of reionisation, 12.6 Gyr ago \citep{2022MNRAS.514...55B}. Furthermore, CVnI is not an isolated case, as SFH studies in LG dSph have found other cases in which galaxies with similar properties have also been able to form stars after the reionisation epoch, such as the previously cited Cetus and Tucana, or Draco and Ursa Minor \citep{2014ApJ...789..147W}, or LGS-3 \citep{2011ApJ...730...14H}.

The UV background acts on the gas in the galactic halos by heating it and eventually preventing it from reaching the density necessary to ignite star formation if the progenitor halo does not reach a critical total mass \citep[e.g][]{2000ApJ...542..535G} strictly above which a dwarf galaxy would be capable of forming stars. In this regard, a recent study by \citet{2023MNRAS.519.1425P} established, through simulations, the redshift evolution of this critical mass boundary through the interplay between it and the mass accretion history of a dwarf’s dark halo, establishing that it would have a value of $\sim$10$^{9.7}$ M$_{\odot}$ at $z$ = 0. This scenario opens up the possibility that the action of the UV background heating could play a significant role in regulating star formation even at ages beyond the end of the reionisation epoch. 

To evaluate this possible effect on the SFH of CVnI, we started by estimating its present-day halo virial mass. To do so, we derived its present-day stellar mass in two different ways. Firstly, following the methodology by \citet{2018MNRAS.479.1514B} and \citet{2021ApJ...909..192G}, we used 2.3$\cdot$10$^{5}$ L$_{\odot}$ as its luminosity \citep{2012AJ....144....4M}. Given that CVnI is a fast galaxy, and according to \citet{2008MNRAS.390.1453W}, we considered that it would present a mass-to-light ratio of $\sim$1.6, which allowed us to derive a stellar mass of 3.68$\pm$0.5$\cdot$10$^{5}$ M$_{\odot}$ at $z$ = 0. Secondly, following \citet{2026ApJ...998...20H}, we calculated the mass converted into stars throughout the whole lifetime of the galaxy, integrating its SFR between 13.7 Gyr (our upper age limit) and 2 Myr (our lower age limit), and applying a correction factor to account for the stellar component not included in the HST/ACS observations, which covered approximately 18\% of the galaxy's mass\footnote{According to a comparison of the number of RGB stars in the ACS field of view with those in the ground-based photometry from the collection of homogeneous photometry by P. B. Stetson, https://www.canfar.net/storage/vault/list/STETSON/homogeneous/}. We then corrected the measurement for the recycling of stellar mass to the ISM, with a factor of 0.284 from Table 2 of \citet{2016MNRAS.455.4183V}, based on the estimated metallicity of CVnI (Z $\approx$ 0.00025; [Z/H] $\approx$ $-$1.8) and the IMF of \citet{1993MNRAS.262..545K}. This
yielded an estimate of 4.40$\pm$1.05$\cdot$10$^{5}$ M$_{\odot}$. Both values are reasonably close (<1$\sigma$) and we report a final value of 4.03$\pm$1.16$\cdot$10$^{5}$ M$_{\odot}$ for the present-day stellar mass of CVnI, derived from the average of both methods. We linked this mass to the halo virial mass using the abundance-matching (AM) relation by \citet{2014ApJ...784L..14B}, with their parameter: M$_{0}$ = 63.1. Thus, we estimated that at present, the halo virial mass of CVnI is 10$^{9.6}$ M$_{\odot}$, slightly below the \citet{2023MNRAS.519.1425P} critical mass threshold for star formation, as might be expected. 

Since the SFH of the galaxy indicates that it would have started forming stars at $\sim$12.6 Gyr, quenching almost completely $\sim$10.9 Gyr ago, we evolved its halo virial mass at both moments ($z$ = 6 and $z$ = 2) and at an intermediate value ($z$ = 4). \citet{2010MNRAS.406.2267F} suggested that the mean and median mass growth rates of haloes of mass M at redshift $z$ are governed by their Equation 2, which relates the evolution of the halo mass to the redshift itself, as well as to the cosmological parameters $\omega_m$ and $\omega_\Lambda$. Thus, we numerically integrated Equation 2 (using $\omega_m$ = 0.25, $\omega_\Lambda$ = 0.75, and $H_0$ = 70 km$\cdot$s$^{-1}$$\cdot$Mpc$^{-1}$), to obtain the mean mass of the halo of CVnI at different values of z. For $z$ = 6, CVnI would have a value of 10$^{8.74}$ M$_{\odot}$, above the critical mass at that redshift; for $z$ = 4, it would have a value of 10$^{9.04}$ M$_{\odot}$, still slightly above the critical threshold; and for $z$ = 2, we would have a value of 10$^{9.36}$ M$_{\odot}$, very similar to the critical mass at that time. These results are in good agreement with the calculated SFH, and offer a scenario in which the galaxy would have been capable of forming stars from ages older than 12 Gyr, with its halo mass evolving very close to the mass threshold, but always above it, until it would have become subcritical around 10.9 - 10.4 Gyr ago. At this point, its star formation would have ceased until the present day. It should be noted that the mass derived from the upper end of the uncertainty would imply that the galaxy could have been above the critical mass threshold up to the present day, that is, able to form stars to the present day, which is not the case. On the other hand, the mass derived from the lower end of the uncertainty range would not affect CVnI’s position relative to the critical mass threshold. Thus, it seems reasonable to consider the mean value obtained for the present-day mass to be reliable and  the proposed evolutionary scenario, along with it.

We note also that in this scenario, it is necessary to take into account other possible processes that could have played a significant role in the mass evolution of the galaxy, intrinsic to the object itself, such as the SN feedback, or local, such as the interaction with large galaxies or with a dense environment, or even the possibility of a patchy reionisation, as suggested in recent studies \citep{2021ApJ...920L..19S, 2023ApJ...959...31K}. We explored the role of SN feedback. The plot on the left of Figure \ref{fig:AMR_MDF_CVnI_Johnson_1} shows that the metallicity range of the CVnI stellar population was established very quickly at its practically definitive values, indicating that enough supernova-type events had to occur in a sufficiently short time interval from the beginning of the main episode to enrich the medium substantially. Thus, we consider whether these explosions could have removed enough gas from the galaxy, leading to the quenching of star formation found in its SFH. To evaluate this question, we estimated the number of SNeII that could have occurred during the main star formation period.

First, we calculated the mass converted into stars during the main period, integrating its SFR
between 13.7 Gyr and the quenching age at 9 Gyr, while applying the same correction factor for the stellar component not included in the HST/ACS observations as the one used to calculate the present-day stellar mass. Thus, we obtained $\sim$5.7$\cdot$10$^{5}$ M$_{\odot}$ as the mass transformed into stars during that period. Assuming that SNe II originate in stars with masses greater than 6.5 M$_{\odot}$, and integrating the IMF used \citep{2001MNRAS.322..231K} under that assumption, we estimated that approximately 9610
SNeII must have been produced. Assuming an energy release of 10$^{51}$ erg for each SNeII event \citep{1999ApJS..123....3L}, and $\sim$1.5 Gyr as the real duration of the main star formation period (as estimated from the mock tests described in Appendix \ref{sec:Robustness}), we finally obtained a luminosity of approximately 2$\times$10$^{38}$ erg s$^{-1}$. If we consider the mass of the gas at the time of the quenching of star formation, which we can obtain from its halo mass at $z$ = 2 by applying the baryonic fraction (f$_{b}$ = 1/6), the comparison of it to the luminosity places the galaxy in the blow-out/mass loss regime, according to the model
of \citet{1999ApJ...513..142M}, which establishes the effects of stellar winds and SN explosions from starbursts with mechanical luminosities ranging from 10$^{37}$ to 10$^{39}$ erg s$^{-1}$ on the ISM of dwarf galaxies with masses of 10$^{6}$ to 10$^{9}$ M$_{\odot}$. Based on hydrodynamic simulations in which the gas mass ($M_g$) and the mechanical luminosity injected by supernovae are varied, they track the evolution of a starburst and classify the result according to whether there is no loss (retention), whether only the metals escape (blow-out), or whether all the gas is expelled (blow-away). 

This would indicate that the loss of gas and metals due to the winds produced by SNe could not have been solely responsible for the quenching of star formation, but their combined action with that of the UV background would be compatible with a sufficient loss of material causing the total mass to fall below the critical threshold. As a result, the remaining gas would be prevented from having enough density to continue star formation.

Finally, we look at the possible influence of environmental effects on the evolution of CVnI. The influence of close encounters with giant galaxies on the evolution of dwarf galaxies is an established theory in evolutionary models \citep{2010AdAst2010E..25M, 2010ApJ...708.1032L}, although it typically requires more than one orbit <50 kpc to transform a disk-shaped dwarf into a dSph \citep{2010ApJ...722.1864M}. \citet{2007ApJ...670..313S} derived, for CVnI a heliocentric systemic, a line-of-sight velocity of 30.9 km s$^{-1}$. The orbits of CVnI  inferred in a recent work by \citet{2022A&A...657A..54B}, indicate that its most recent pericentric passage was about 1 Gyr ago \citep[see Table 5 in][]{2022A&A...657A..54B}. Depending on the gravitational potential adopted for the MW, the orbits are compatible with the galaxy having experienced other pericentric passages at early times (around 5 Gyr ago and 9 Gyr ago). It is interesting to note that these possible early close encounters with a giant galaxy is another similarity between CVnI and Tucana and Cetus, other clear outliers of the morphology-density relation. On the other hand, another important effect for removing gas in close encounters with giant galaxies is ram pressure stripping, which should remove the gas fast, once the galaxy has entered the MW halo. Thus, although it is not possible to estimate the effect of the environment in the evolution of CVnI with absolute certainty, it cannot be completely ruled out that a possible passage close to the MW acted in a complementary way to the SNeII in the expulsion of gas and in the quenching of star formation as a result.

\section{Summary and final remarks}\label{sec:Summary}

We studied the stellar evolution of CVnI, a distant dSph satellite of the MW. Using archival HST/ACS observations of two adjacent fields located to the north and east of the centre of CVnI, we obtained the deepest CMD to date. Our approach allowed us to derive the SFH of the galaxy using the most up-to-date BaSTI-IAC stellar evolution libraries that include diffusion, together with state-of-the-art CMD fitting techniques, such as ChronoSynth \citep{2024A&A...687A.168G}, DisPar \citep{2021MNRAS.501.3962R} and DirSFH \citep{2024A&A...687A.168G}.

In the treatment of the photometric data, the usual artificial star test techniques widely contrasted in the literature were used to adequately characterise the observational effects and simulate them to the synthetic data used in the calculation of the SFH. Likewise, exhaustive tests have been carried out to verify the consistency of the SFH calculation against changes in the heavy element distribution of the stellar models or the binary star fraction. We also estimated  our temporal resolution by obtaining the SFH of mock populations and a globular cluster, M92, whose properties are well established in the literature. 

We found that the SFH of CVnI is dominated by an old episode of star formation, with its peak at around 11.8 Gyr and whose temporal duration can be estimated to be between 1 and 1.5 Gyr, followed by an apparently short and very mild period centred on 8 Gyr. This old stellar population is mainly metal-poor, with the metallicity ranging from [Z/H] $\sim$ $-$2.4 dex to [Z/H] $\sim$ $-$1.5 dex, and a peak around $\sim$ $-$1.8 dex. This is in good in agreement with published spectroscopic metallicity for RGB stars. Moreover, our study excludes the presence of stars younger than 8 Gyr or more metallic stars. Interestingly, in the surveyed central regions, we did not detect any spatial gradients in terms of age or metallicity distribution.

Remarkably, the peak of star formation in CVnI 11.8 Gyr ago occurred after the reionisation epoch, assumed to end around 12.6 Gyr ago. While it could be argued that this difference may be attributed to a systematic offset in the age zero-point, the fact that the peak age we derived for M92 is 12.9 Gyr supports the conclusion that this is a truly delayed time for this peak, as compared to the general onset of star formation in the Universe. Compared with CVnI, other fast dSph satellites of the MW with similar luminosity or size, such as Draco, Ursa Minor or Sextans, exhibit older and shorter lived SFHs, \citep[see][and references therein]{2015ApJ...811L..18G}. This implies that CVnI would be one of the youngest fast MW satellites. Given that all these galaxies are much closer to the MW, it would not be unreasonable to assume that the distance from it may have played a non-negligible influence on the delayed star formation peak of CVnI, as is the case with the SFHs of other also distant fast galaxies (although much more luminous), such as Cetus or LGS-3 \citep[see][and references therein]{2015ApJ...811L..18G}. We analysed the main mechanisms that may affect the ignition and intensity of star formation in dwarf galaxies and reached the following conclusions:

Regarding the impact of the UV background in modulating star formation, we estimated the evolution of the virial mass of its halo according to the models of \citet{2010MNRAS.406.2267F} and compared it with the critical mass thresholds for star formation at different $z$ derived by \citet{2023MNRAS.519.1425P} using the APOSTLE simulations. We found the virial masses of CVnI to be slightly above the critical thresholds, in agreement with the fact that CVnI has been forming stars through the reionisation epoch. We also considered the effects of SNeII in removing gas from the galaxy. The mass transformed into stars during the main star formation period, together with the estimate of the energy released by SNeII, places the galaxy in a regime of blow-out/mass loss according to the model of \citet{1999ApJ...513..142M}. Thus, SNe are not expected to be solely responsible for the quenching of the star formation. Finally, regarding environmental effects, published orbital estimates for this galaxy do not exclude the possibility that CVnI might have passed close to (or even crossed) the MW halo at ages similar to the quenching of star formation. However, the role played by interactions in the evolution of the galaxy remains uncertain, since it is not possible to establish a specific age for this close passage with the data currently available. 

The results obtained and the analyses we carried out in this work allow us to outline a possible scenario for the evolution of CVnI: at very early ages, close to the end of the reionisation epoch, the galaxy was able to accrete enough mass to exceed the critical mass threshold and ignite star formation. From then on, the evolution of the galaxy's accretion history caused the mass of its halo to evolve above, but very close to the critical threshold; given this proximity to the threshold value, star formation would have been very sensitive to possible mass losses. From the beginning of this star formation period (and presumably throughout its duration), a sufficient number of supernova events were generated to increase the metallicity of the medium in a very short period of time to its final values and to gradually expel gas. This gradual gas loss would have led to the total mass of the galaxy becoming subcritical around 10.4 Gyr ($z$ = 2), resulting in the definitive quenching of star formation. It cannot be ruled out that tidal effects may also have contributed to this loss of mass due to a possible close passage of the galaxy to the MW halo at an age close to the time of the quenching. However, this same interaction could explain the small star formation period found 8 Gyr ago, where a possible accretion of material could have caused the total mass to temporarily exceed the critical threshold again.

The HST/ACS data used to calculate the SFH are localised in two central and relatively small areas of the galaxy. \citet{2024ApJ...977...20R} suggested that dwarf galaxies could present age gradients that show older stars tend to be displaced towards the outer regions. If this were the case in CVnI, it could be that the applied data only referred to the younger part of the age range actually present in it, giving rise to an apparent delayed time of its star formation peak. Therefore, it would be crucial to recalculate the SFH with data from outer regions. New surveys, such as the future Nancy Roman Space Telescope \footnote{https://science.nasa.gov/mission/roman-space-telescope/}, or upcoming data releases from the Euclid mission \footnote{https://sci.esa.int/web/euclid} could help provide sufficient new information from other areas of the galaxy that would prove suitable for expanding the study of the SFH.

\begin{acknowledgements}
We would like to thank Massimo Marengo for the interesting discussions we have had about these results in the Gaia RRL Zoom meetings. Also we would like to thank Alan McConacchie, Kathy Vivas and Antonela Monachesi for their useful suggestions in Florence and Cordoba. P.P-G and EFA acknowledge support from the Agencia Estatal de Investigación del Ministerio de Ciencia e Innovación (AEI-MCINN) under grant 'At the forefront of Galactic Archaeology in the Gaia era: the chemical aspects' with reference PID2023-150319NB-C22/10.13039/501100011033. M.Mo, CG and GB acknowledge support from the Agencia Estatal de Investigaci\'on del Ministerio de Ciencia e Innovaci\'on (AEI-MCINN) under grants 'At the forefront of Galactic Archaeology: evolution of the luminous and dark matter components of the MW and LG dwarf galaxies in the {\it Gaia} era' with reference PID2023-150319NB-C21/10.13039/501100011033. M.Mo. acknowledges support from the Spanish Ministry of Science, Innovation and Universities (MICIU) through the Spanish State Research Agency under the grants 'RR Lyrae stars, a lighthouse to distant galaxies and early galaxy evolution' and the European Regional Development Fun (ERDF) with reference PID2021-127042OB-I00. CG and M. Mo. also acknowledge support from the Severo Ochoa Program 2020-2023 (CEX2019-000920-S). LC thanks the support from grant IA103326 (DGAPA-PAPIIT, UNAM) and project SECIHTI CBF-2025-I-2048. LC acknowledges support from the Fundación Occident and the Instituto de Astrofísica de Canarias under the Visiting Researcher Programme 2022–2024 agreed between both institutions. CG acknowledges financial support from the European Research Council (ERC-2024-AdG "ChronoGal: Chronology of our Galaxy from Gaia CMD-fitting", Grant Agreement 101201412). Views and opinions expressed are, however, those of the author(s) only and do not necessarily reflect those of the European Union or the  European Research Council Executive Agency (ERCEA). Neither the European Union nor the granting authority can be held responsible for them. The author(s) wish to acknowledge the contribution of the IAC High-Performance Computing support team and hardware facilities to the results of this research. This paper made use of the IAC HTCondor facility (https://research.cs.wisc.edu/htcondor/), partly financed by the Ministry of Economy and Competitiveness with FEDER funds, code
IACA13-3E-2493. 

\end{acknowledgements}

\bibliographystyle{aa} 
\bibliography{PedroJPerez_bibtex.bib}

\begin{appendix}

\section{Changes in the SFH resulting from removing possible non-CVnI stars from the CMD.}\label{sec:extra_figures}

\begin{figure}[H]
\centering
\includegraphics[width=\columnwidth] { 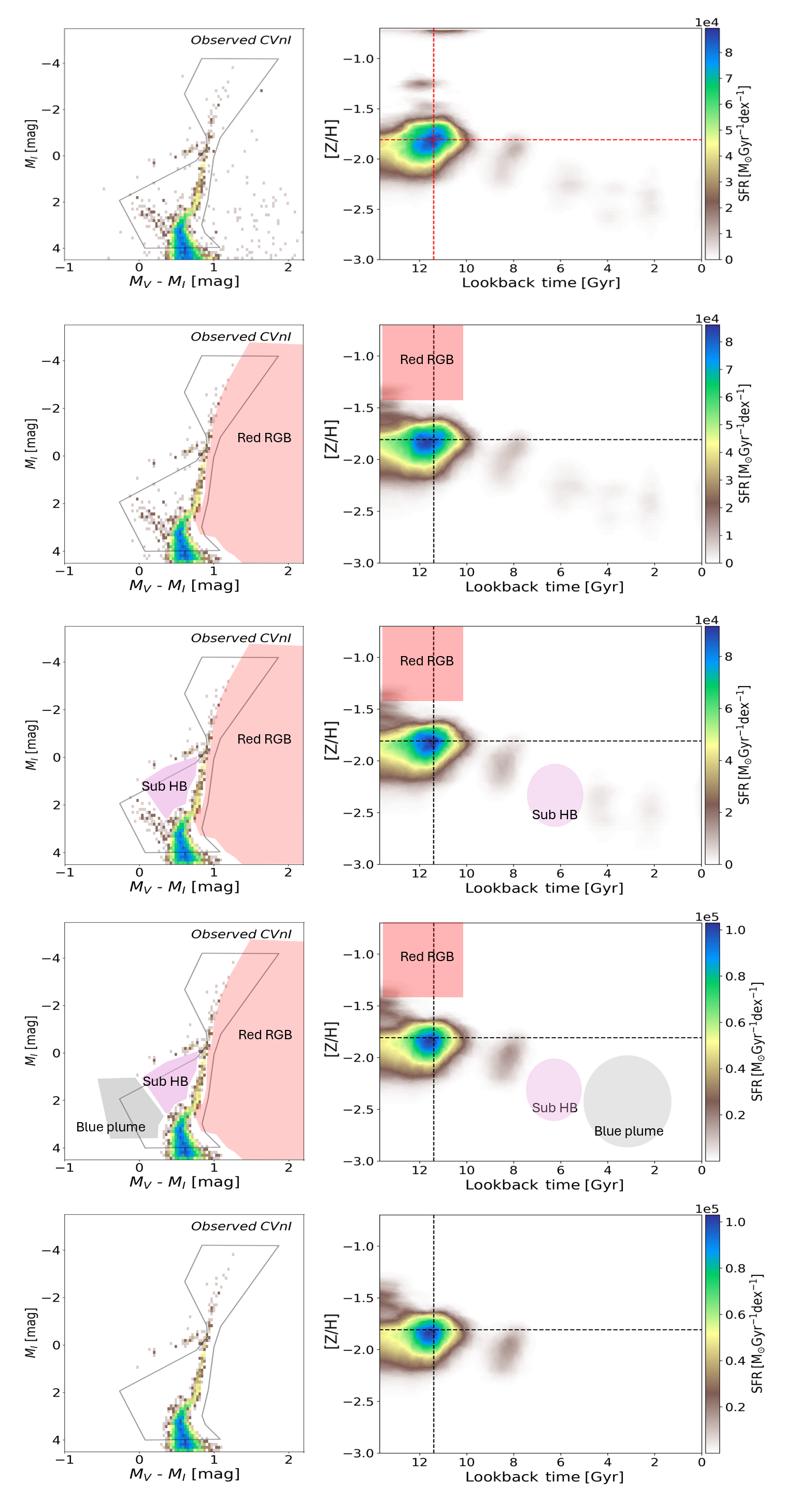} \\
\caption{Results of tests consisting of progressively cleaning the CVnI CMD. The coloured areas mark the subtracted stars in the CMD and the related star population that is no longer found in the solution SFH. See text for full details.}  
\label{fig:CMD_NoRGB_NosubHB_noBP} 
\end{figure}

\begin{figure}
\centering 
\includegraphics[width=0.7\columnwidth,height=1.5\columnwidth]{ 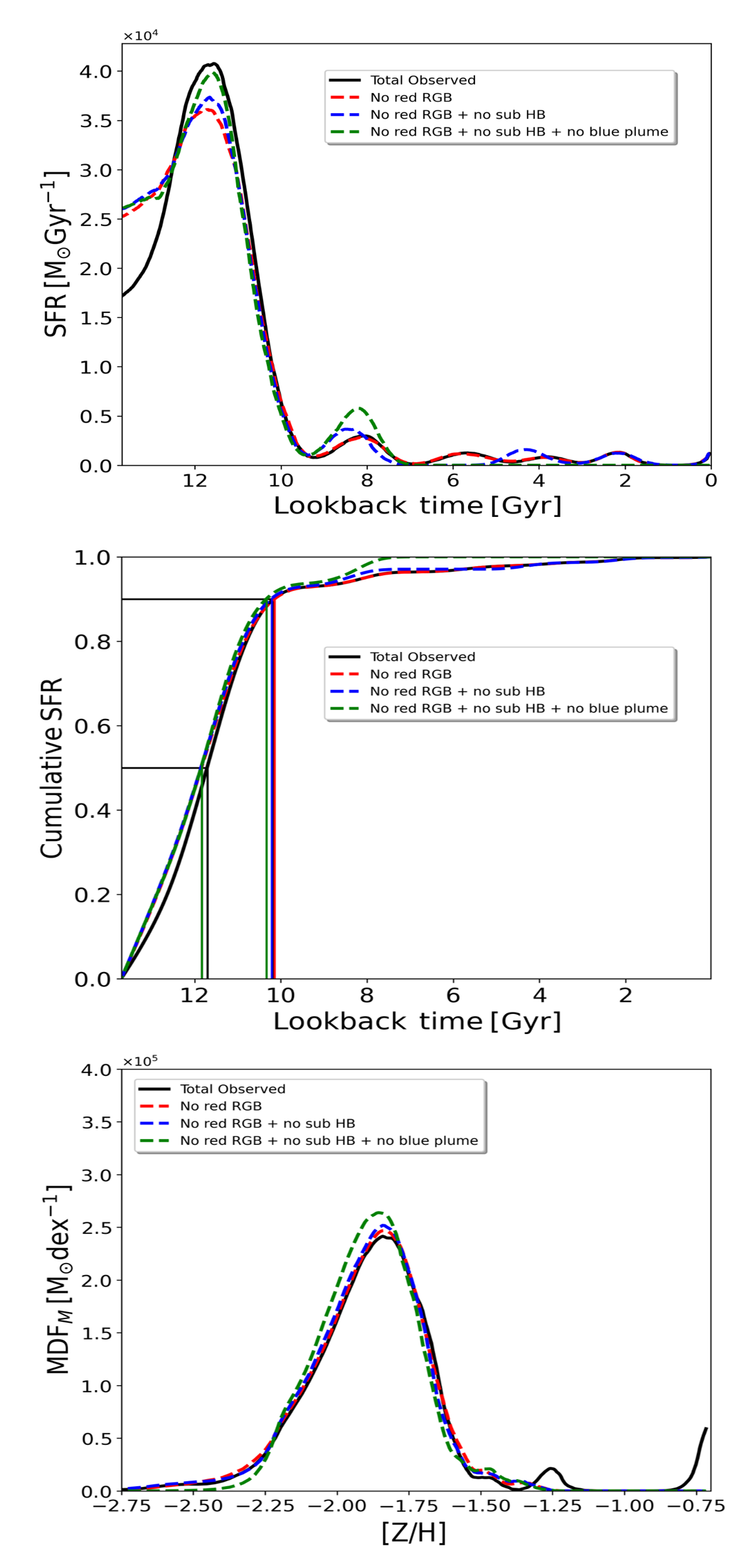} \\
\caption{Comparison of the SFR, cumulative SFR, and MDF curves of the CVnI SFH after progressively cleaning the observed CMD. See text for full details} 
\label{fig:SFH_NoRGB_NosubHB_noBP} 
\end{figure}

As cited in Section \ref{sec:bss}, we performed a set of tests to try to clarify the apparent populations of metal-rich old stars and metal-poor young stars found in the SFH. We sequentially subtracted stars from specific areas of the original CMD and calculated the SFH of these progressively 'cleaned' CMDs. Figure \ref{fig:CMD_NoRGB_NosubHB_noBP} shows the results of these tests in the AMR plane. The first row shows the original CVnI CMD and AMR, while the lower rows show, from top to bottom, the effects of the sequential subtractions: The second row show the SFH solution after subtracting the stars redder than the RGB, and it can be seen that the effect in the SFH is the disappearance of the apparently old ($\sim$ 12 Gyr) and very metal-rich populations; the third row shows the SFH after the subtraction of the stars located between the HB and the SGB, which effect in the SFH is the disappearance of the apparent population at 6 Gyr; the fourth row shows the SFH after the final subtraction of the blue plume stars from the CVnI CMD, resulting in the complete disappearance of the apparent populations younger than 6 Gyr. Finally, the last row shows the same as the fourth row without the coloured areas, i.e. the completely clean CMD and the calculated effect on the AMR.

On the other hand, Figure \ref{fig:SFH_NoRGB_NosubHB_noBP} shows the comparison between the SFR, cumulative SFR and MDF of the SFH solutions of each test. The black lines in the three plots are the curves obtained for the original solution calculated with all ACS observed stars. The dashed lines are those obtained with the progressive subtractions of stars from the original CMD. These plots show that the subtractions do not affect the overall SFH obtained for CVnI, which consistently maintains its main characteristics. The subtraction of stars has a small effect on the height of the SFR and MDF curves, due to the small number of stars discarded, and on the cumulative curve, shifting it slightly towards older ages, due to the removal of apparently young stars.

\section{Robustness of the solution}\label{sec:Robustness}

\begin{figure}
\raggedright
\includegraphics[width=\columnwidth]{ 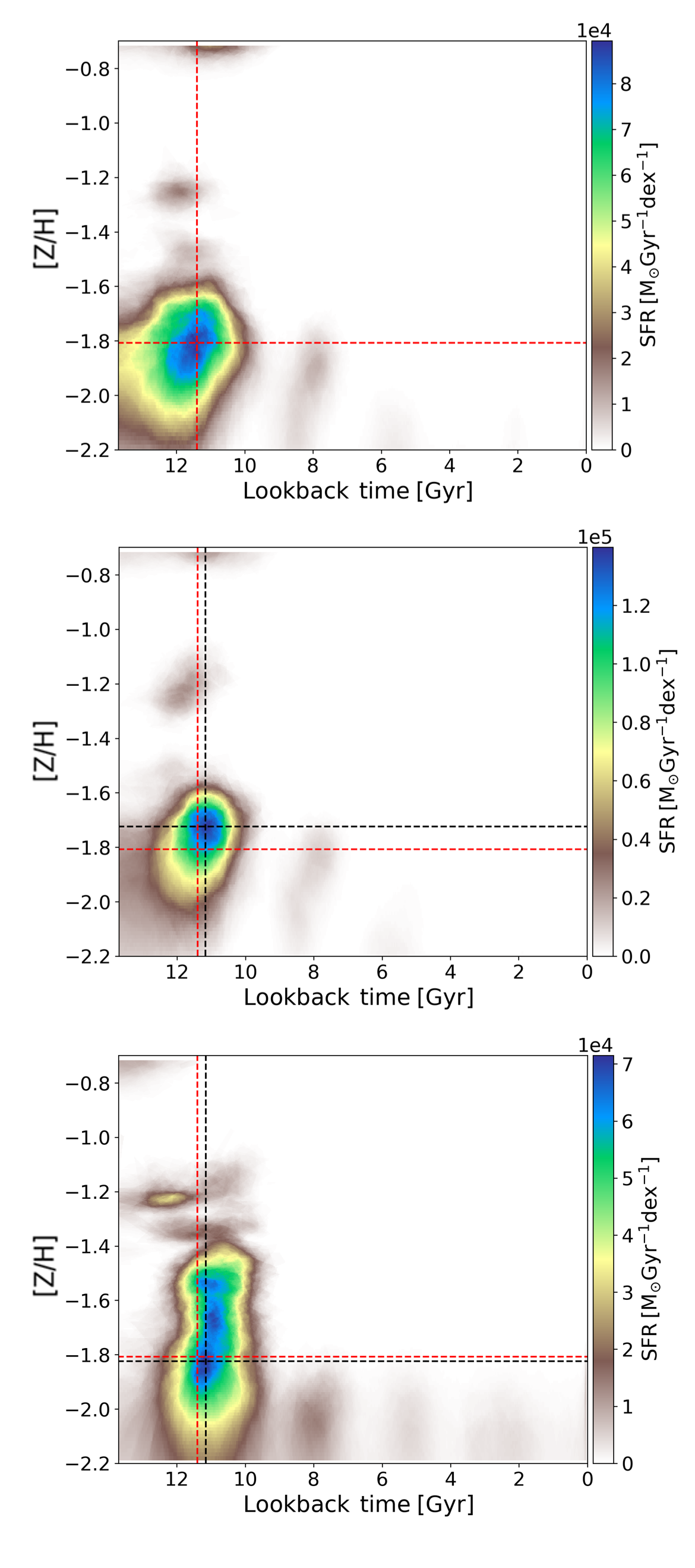} \\   
\caption{Comparative of the AMR solutions obtained with the BaSTI solar-scaled and $\alpha$-enhanced mother CMDs, and the PARSEC solar-scaled mother CMD. Top: AMR solution with BaSTI solar-scaled mother CMD, the same already shown in Figure \ref{fig:AMR_MDF_CVnI_Johnson_1}. Centre: AMR solution with BaSTI $\alpha$-enhanced mother CMD. Bottom: AMR solution with PARSEC solar-scaled mother CMD. In all solutions the dashed red lines mark the location of the star formation peak for the BaSTI solar-scaled solution, while the dashed black lines in the centre and bottom plots mark the location of the peaks for those solutions. The similar location of the peaks in the three cases shows the robustness of the results (the scale of the [Z/H] axis has been adapted to the different lower limit in Z of the PARSEC mother CMD).}
\label{fig:CVnI_SFH_Comparative_Three_solutions} 
\end{figure}

\begin{figure*}
\centering
\includegraphics[width=\textwidth]{ 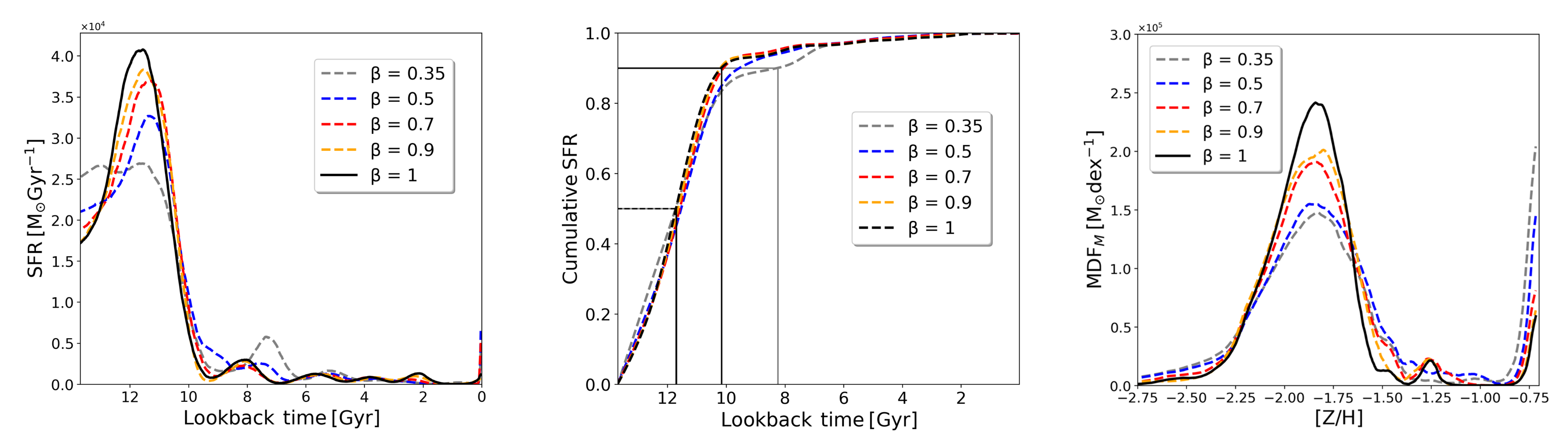} \\   
\caption{Comparison of the SFR (left), Cumulative SFR (centre) and MDF (right) solutions obtained using the solar-scaled mother CMDs with different Binary fractions.}  
\label{fig:CVnI_SFH_Comparative_Binarity_solutions} 
\end{figure*}

The CMD-fitting technique is undoubtedly considered reliable and robust for studying the stellar evolution of a galaxy, but it is still subject to a number of variables that can lead to uncertainties in the results. The choice of initial parameters such as, for example, the range and width of the bins in age and metallicity, the shape of the bundle, the specific characteristics of the synthetic CMDs or even the position of the best shifts during the calculation itself, can have non-negligible effects on the solutions obtained. Taking this into account, we designed and executed a series of tests aimed at ensuring the accuracy and resolution of the results obtained.  The tests consisted of:

\begin{itemize}

\item Test the solutions using different stellar libraries to generate the mother synthetic CMDs.
\item Use different fraction of binary stars when generating the mother synthetic CMDs. 
\item Use synthetic populations with well-defined ages and metallicities to verify the accuracy and resolution of the solutions.  
\item Finally, use a real object with an stellar population of well-known age and metallicity, to check the accuracy and correctness with which the process reflects them in its solution. 
\end{itemize}

\subsection{Effect of the stellar library}\label{sec:library}

The use of a specific stellar library may be subject to systematic uncertainties or biases. To assess whether this might be occurring when using the solar-scaled BaSTI library to obtain the SFH of CVnI, we also calculated it using two other stellar libraries: an $\alpha$-enhanced BaSTI library and a solar-scaled PARSEC library. Figure \ref{fig:CVnI_SFH_Comparative_Three_solutions} compares the AMR of the solutions obtained with BaSTI solar-scaled and $\alpha$-enhanced mother CMDs (top and centre, respectively), maintaining the same parameters listed in Section \ref{sec:cmdsynth}, and with PARSEC solar-scaled mother CMD (bottom), generated with a minimum Z value of 0.0001 ([Z/H] = $-$2.2) due to the existing lower limit in the library. The plots show substantial agreement in the age ranges of star formation and in the overall metallicity of all structures (the most noticeable difference is seen in the metallicity distribution of the PARSEC solution, which extends approximately 0.2 dex further towards higher values of [Z/H]). Nevertheless, note also the similarity in the position of the star formation maxima in the three cases, which differ by at most $\sim$ 0.2 Gyr in age and $\sim$ 0.1 dex in [Z/H]. This agreement across the comparisons confirms that the derived SFH is robust. 

\subsection{Effect of the fraction of binary stars}\label{sec:binary}

Similarly, we checked the effect of changing the binary fraction in the mother CMD. We generate additional mother CMDs, with $\beta$ = 0.35, $\beta$ = 0.5, $\beta$ = 0.7, and $\beta$ = 0.9, respectively, and we recover the SFH with the same parametrisation as the original solution (maintaining the same q) in all cases. The comparative results are shown in Figure \ref{fig:CVnI_SFH_Comparative_Binarity_solutions}. It is remarkable that the change in the binary fraction, which is an unknown parameter that must be assumed in the calculation of the synthetic mother CMD, only reflects appreciable differences in the gradual and slight decrease in the mass converted into stars as the binary fraction decreases (less than 7\% between $\beta$ = 1 and $\beta$ = 0.35), which is a reasonable effect, and in the greater intensity of the star formation period at 8 Gyr revealed with $\beta$ = 0.35. None of these differences significantly affects the position of the SFR or MDF peaks, nor the extent of star formation or the width of the metallicity distribution.

\subsection{Accuracy and precision of the results. Mock populations}\label{sec:Mock_pops}

\begin{figure} 
\centering 
\includegraphics[width=\columnwidth]{ 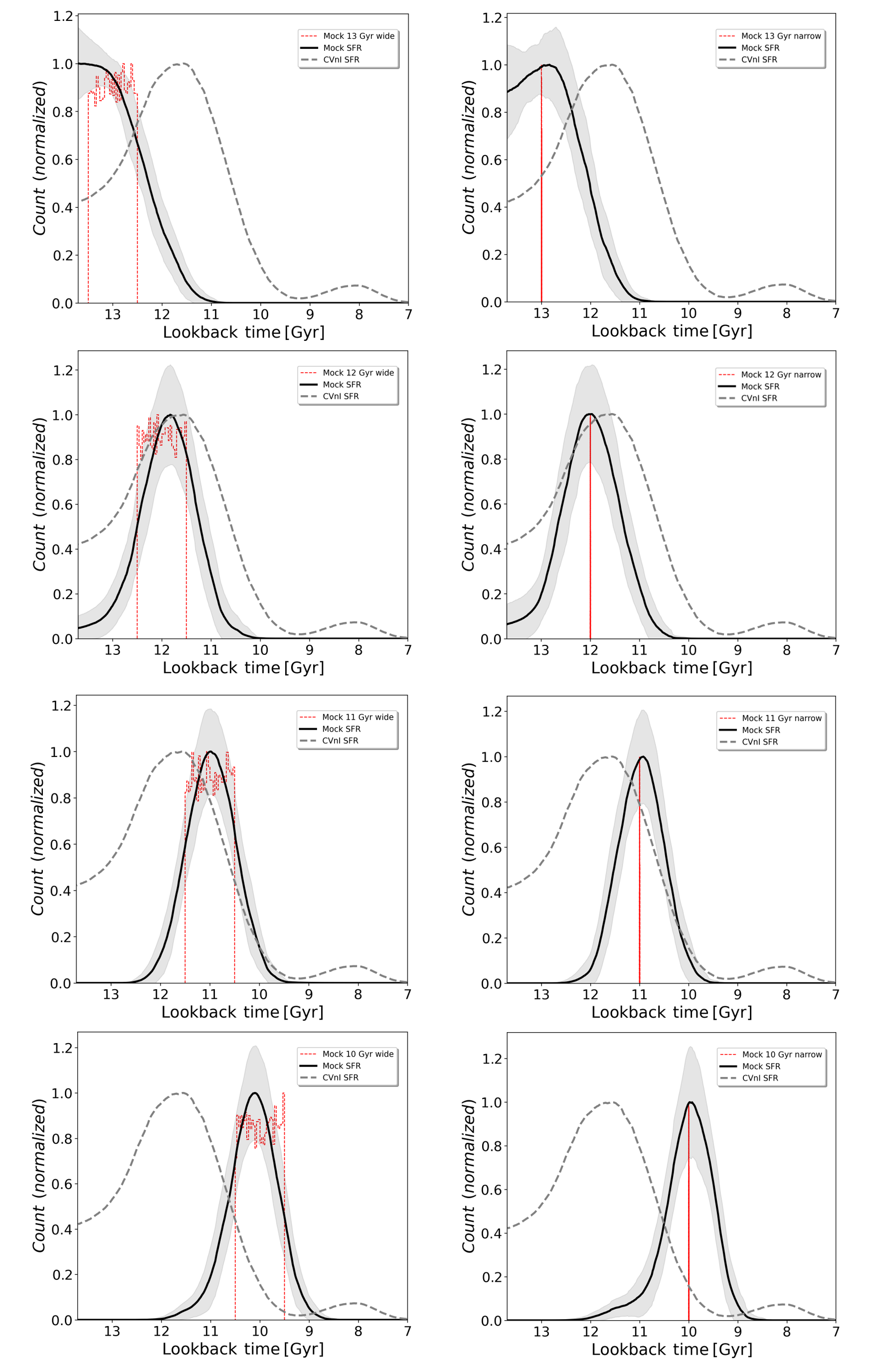} \\
\caption{SFR curves of the eight mock populations compared to the original age duration of each of them, and to the SFR of CVnI. See text for full explanation} 
\label{fig:Mock_SFR_curves} 
\end{figure}

\begin{figure} 
\centering 
\includegraphics[width=\columnwidth]{ 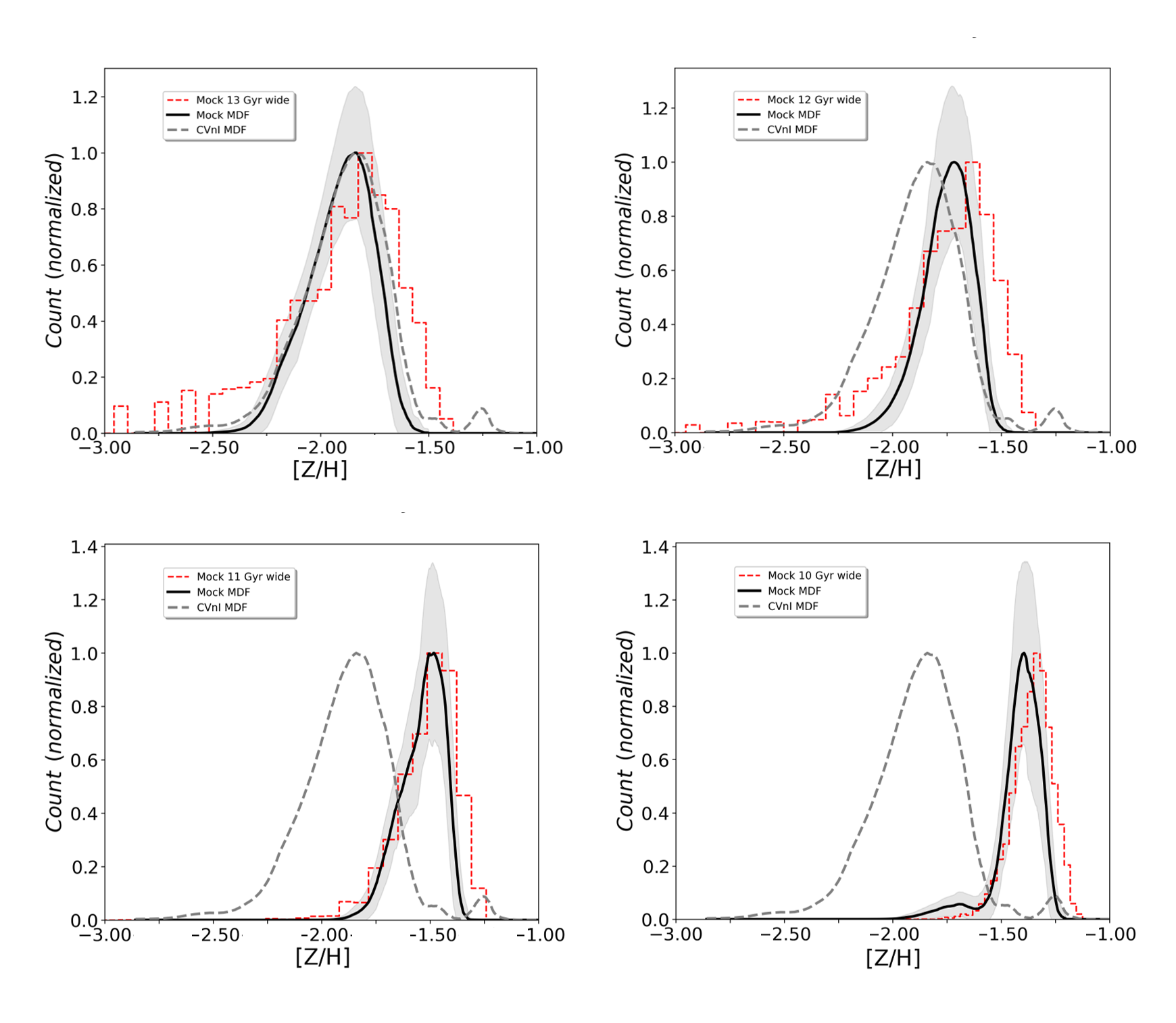} \\
\caption{MDF curves of the four 1 Gyr-wide mock populations compared to the original metallicity range of each of them, and to the MDF of CVnI. See text for full explanation.} 
\label{fig:Mock_MDF_curves} 
\end{figure}

dirSFH has allowed us to obtain results with unprecedented precision in the SFR and the metallicity distribution of the population of CVnI. Nevertheless, the measured width of both the star formation episodes and the MDF is known to be artificially increased due to the finite intrinsic resolution of the method, which is driven, among others, by the finite size of the SSPs, the photometric errors and the small separation of the stellar populations of different age and metallicity in the CMD, specially at old ages. It is therefore necessary to estimate the actual resolution or precision of our solutions to provide a better estimate of the actual characteristics of the stellar populations in CVnI. Additionally, systematics in the determination of the actual age and metallicity may occur, and thus the accuracy of the solutions also needs to be estimated. The purpose of this section is to show how we estimated both the accuracy and the precision of our SFHs by means of recovering mock stellar population of known age and metallicity distributions. 

Using the same BaSTI libraries we worked with for the rest of the study, we created eight mock populations with age and metallicity characteristics described in Table \ref{tab:SFR_mocks} of the text. 

The observational errors were simulated using DisPar in the eight mock populations, and by means of dirSFH, their SFH has been calculated using the same age and metallicity bins as for the original CVnI calculation. Figures \ref{fig:Mock_SFR_curves} and \ref{fig:Mock_MDF_curves} show the results obtained for the SFR and the MDF of the eight mock populations (black solid curves). Each curve is superimposed on the actual age and metallicity distributions, respectively, of the mock population (red dotted lines). Equally, the computed SFR and MDF calculated for CVnI in Section \ref{sec:Result_CVnI_SFH} (dashed curves) are shown to compare to the mock solutions.  

Each SFR solution has been fitted with a Gaussian, and their centre and width are shown in Table \ref{tab:SFR_mocks}. The mean ages of the populations are systematically well recovered, with a difference of less than 0.2 Gyr being found in all cases except for the 13 Gyr-Wide population, for which the difference is 0.35 Gyr. Note, however, that in this case the difference is within the errors (indicated by the shadowed area around the solution curve). The mean recovered ages tend to be found slightly younger than the mean input age, except for the 13 and 10 Gyr-Wide mocks. It is also observed that the width of the calculated star formation periods is of the order of 1 Gyr from 12 Gyr onwards and tends to decrease towards younger ages. Moreover, the widths of the periods are very similar for each of the two populations of the same age, indicating that we are getting a time resolution of about 1 Gyr, which is a reasonable result. So, the process would not be able to separate the possible existence of shorter bursts, or to distinguish between a very narrow burst or a star formation episode of $\simeq$ 1 Gyr duration. Comparison with the SFR of CVnI shows that it could be reasonably well represented by the 12 Gyr (mainly) and 11 Gyr mocks, except for the tail present at ages older than 12.5 Gyr. This could indicate that the star formation in the galaxy was increasing with time, beginning at very early ages and reaching its peak around 11.8 Gyr. The comparison allows us to estimate the duration of the main star formation period at a maximum of $\sim$1.7 Gyr, based on the FWHM of the solutions obtained for CVnI itself and the 12 and 11 Gyr mocks. 

Regarding the MDFs, Table \ref{tab:SFR_mocks} summarises the characteristics of the solutions for the populations generated with a stellar formation width of 1 Gyr. The precision in obtaining the mean metallicity values is found to be excellent, with differences of less than 0.1 dex and within the margins of error in all cases (the shadowed areas around the solution curves) in Figure \ref{fig:Mock_MDF_curves}.  It can be seen that the obtained MDF for CVnI fits almost exactly that of the mock population with [Z/H] = $-$ 1.83, indicating that the stellar population of the galaxy could be well represented by this metallicity range. Note that the metallicity range of the solutions is slightly, but systematically, narrower than the real populations. Therefore, it could also be assumed that the real metallicity range of CVnI could be somewhat wider towards higher metallicities than those represented by the solution. 

\begin{figure*} 
\sidecaption
\includegraphics[width=12cm] {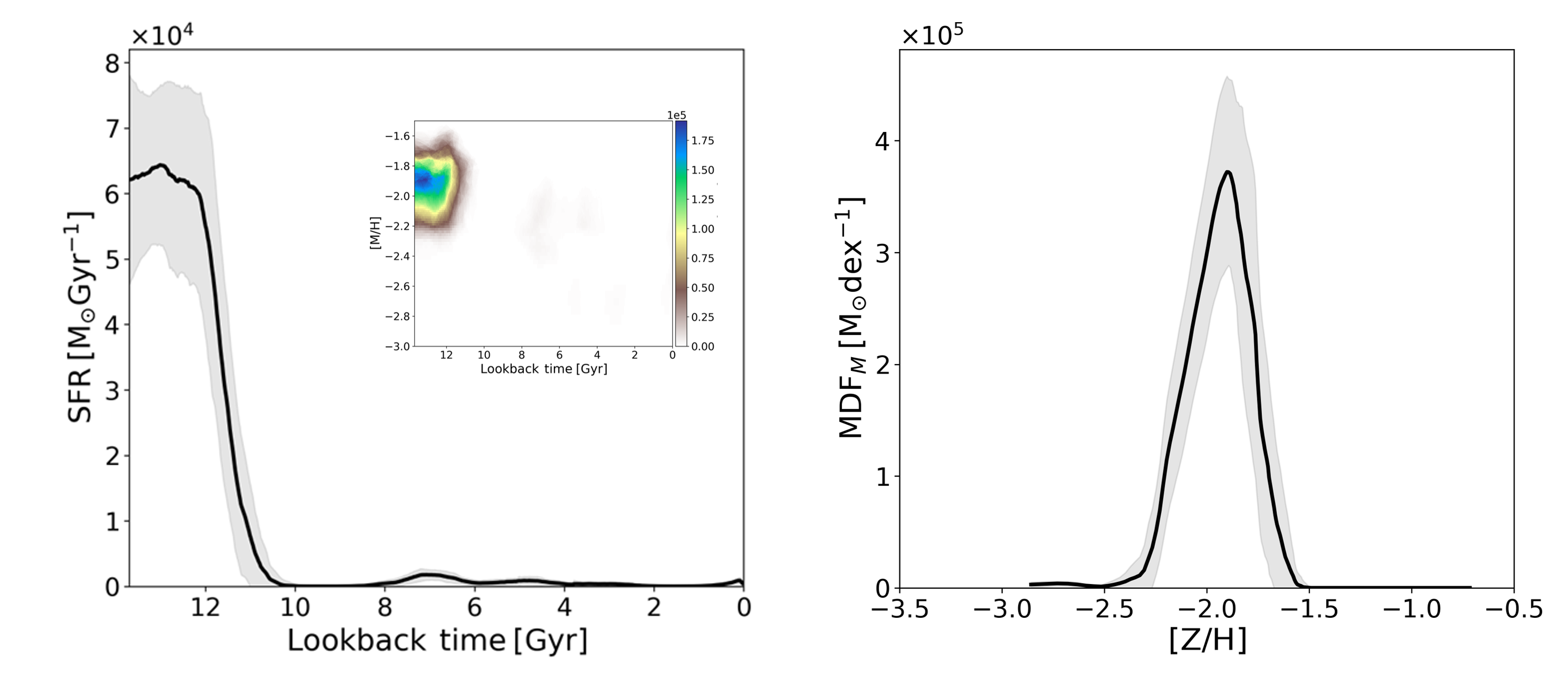}   
    \caption{SFR and MDF curves of M92 obtained with the same calculation process and mother CMD used for CVnI. The age-metallicity map is inserted inside the SFR plot} 
    \label{fig:SFH_M92} 
\end{figure*}

\subsection{M92}\label{sec:M92}

Both the CVnI SFH solutions described in Section \ref{sec:Result_CVnI_SFH} and the test results discussed in the previous subsections were performed using synthetic CMDs and simulated populations generated from stellar evolution models, so it could be argued that they could be affected by some kind of bias due to the use of too ‘ideal’ synthetic populations. 

Therefore, a specific test has been designed using a widely studied real object, whose absolute age and metallicity characteristics can be directly compared with the solution obtained for its SFH. Thus, the well-studied globular cluster M92 (NGC6341) has been chosen, whose published age, between 12 and 14 Gyr, and metallicity, in a [Fe/H] range of  $-$2.3 to $-$2.6 \citep{2010arXiv1012.3224H, 2010PASP..122..991D, 2016ApJ...827....2V, 2023AJ....166...18Y,2025A&A...704A.255A}, can serve as a good touchstone to evaluate the accuracy of the SFH calculation methodology used in this work.

The same observational data as those used in \citet{2021ApJ...909..192G}, corresponding to the HST programme ID 10505, were used. This data were de-reddened with the same colour excess and distance modulus values (E(B-V) = 0.01 $\pm$ 0.01 mag and $\mu$$_{0}$ = 14.59 $\pm$ 0.02 mag) used by \citet{2025A&A...704A.255A}. To perform this test under conditions most similar to the SFH calculations done for CVnI, we selected a subset of stars of similar size and, likewise, the same bundle, and bins in age and metallicity were used. Considering that the M92 photometric errors are negligible compared to those of CVnI, following a reasoning similar to that of \citep{2021ApJ...909..192G}, the CVnI observational errors were reproduced on the M92 data. 

The result of the calculation for M92 is shown in Figure \ref{fig:SFH_M92}. The SFR and MDF have Gaussian fitted peaks at Age = 12.9 with $\sigma$ =  1 Gyr, and [Z/H] = $-$1.9 with $\sigma$ = 0.15, respectively. To transform this global metallicity [Z/H] to [Fe/H] we used Equation \ref{eq:metalicidad} cited in Section \ref{sec:SFH_vs_lit}. In the case of M92, adopting an $\alpha$-enhanced composition with [$\alpha$/Fe]= 0.4, as generally assumed in the literature \citep[see, e.g.][]{2022MNRAS.509.4208V}, the value of [Z/H] = -1.9 becomes [Fe/H] $\sim$ $-$2.2 (< 1$\sigma$ of the known [Fe/H] range for the cluster).  That is, both the SFR and the MDF obtained are in good agreement with the values found in the literature for the absolute age and metallicity of the GC. These results help to ensure that the results found for CVnI are not affected by an undesirable effect or by some unknown bias in the calculation process used.

\end{appendix}

\end{document}